\documentclass[twocolumn,preprintnumbers,amsmath,amssymb,showpacs,superscriptaddress,prl,aps,10pt]{revtex4-1}

\usepackage{graphicx}
\usepackage{bm}
\usepackage{dcolumn}
\usepackage{amssymb}
\usepackage{amsmath}
\usepackage{amsfonts}
\usepackage{epsfig}
\usepackage{graphicx}
\usepackage{stackrel}
\usepackage{paralist}
\usepackage[rgb]{xcolor}
\usepackage{subcaption}
\usepackage{todonotes}
\usepackage{pdfcomment}

\newcommand{\I}{\mathrm{i}}
\newcommand{\refEq}[1]{Eq.~(\ref{#1})}
\newcommand{\refFig}[1]{Fig.~\ref{#1}}

\newcommand{\refApp}[1]{Appendix~}
\newcommand{\citeRef}[1]{Ref.~[\onlinecite{#1}]}

\begin{document}
\title{Gapless Topological Superconductors - Model Hamiltonian and Realization}
\author{Yuval Baum}
\affiliation{Department of Condensed Matter Physics, Weizmann Institute of Science, Rehovot 76100, Israel}
\author{Thore Posske}
\affiliation{Institut f{\"u}r Theoretische Physik und Astrophysik, Universit{\"a}t W{\"u}rzburg, 97074 W{\"u}rzburg, Germany}
\author{Ion Cosma Fulga}
\affiliation{Department of Condensed Matter Physics, Weizmann Institute of Science, Rehovot 76100, Israel}
\author{Bj{\"o}rn Trauzettel}
\affiliation{Institut f{\"u}r Theoretische Physik und Astrophysik, Universit{\"a}t W{\"u}rzburg, 97074 W{\"u}rzburg, Germany}
\author{Ady Stern}
\affiliation{Department of Condensed Matter Physics, Weizmann Institute of Science, Rehovot 76100, Israel}

\begin{abstract}
The existence of an excitation gap in the bulk spectrum is one of the most prominent fingerprints of topological phases of matter.
In this paper, we propose a family of two dimensional Hamiltonians that yield an unusual class $D$ topological superconductor with a gapless bulk spectrum but well-localized Majorana edge states.
We perform a numerical analysis for a representative model of this phase and suggest a concrete physical realization by analyzing the effect of magnetic impurities on the surface of strong topological insulators.
\end{abstract}
\maketitle

\textit{Introduction ---}
The pursuit of new topological phases of matter has lead to the discovery of novel quantum states and exotic excitations \cite{Hasan2010,FuandKane,nonabelian}.
The topological classification of phases relies on the existence of an excitation gap in the bulk spectrum. Nonetheless, exceptions exist, where a system may exhibit topological behavior also in the absence of such a bulk gap \cite{Majoranaflatbands,Keselman}. Among these exceptions are, e.g., Weyl semi-metal and nodal superconductors \cite{Schnyder2013,Queiroz}, which possess topologically stable Fermi points or nodal lines respectively.
Additionally, it was shown in \citeRef{ourpaper} that the simultaneous existence of one-dimensional gapless modes on the edge and gapless modes in the two-dimensional bulk may arise in a family of two-dimensional Hamiltonians, generated by coupling a topological phase to a gapless phase.

In this paper, we present a scheme of generating an intrinsic gapless superconducting phase in symmetry class D \cite{class,Kitaev} with well localized edge states. 
This phase is distinct from the previous proposals. First, it emerges
from intrinsic degrees of freedom as opposed to relying on an engineered coupling between topological phases and gapless ones. Second,
it may spontaneously form on the surface of three dimensional topological insulators, which provides a new experimental route for realizing and probing gapless topological phases. Finally, the chirality of its edge modes depends on disorder in an unusual manner.
We suggest a concrete example of this phase: a Rashba two-dimensional electron gas (2DEG) in the presence of a
modulated magnetization that is proximity coupled to an s-wave superconductor. We analyze the spectral and the
transport properties of this model both for clean systems and in the presence of disorder.

In \citeRef{ourpaper}, we have found that in the clean case Dirac excitations in the bulk coexist with two types of edge states, depending on the edge orientation. For most orientations, we have identified ``strong'' edge states that do not hybridize with the bulk states due to energy and momentum conservation. Their wave functions remain exponentially localized near the edge despite the absence of a bulk gap. For some orientations, 
this is not the case, and the edge states wave functions leak into the bulk, making the edge modes ``weak''. Unlike in the previous scheme, here, the intrinsic origin of the edge states leads to an additional and surprising dependence of the edge states chirality on the lattice termination. 

\begin{figure}
\centering
\includegraphics[width = \linewidth]{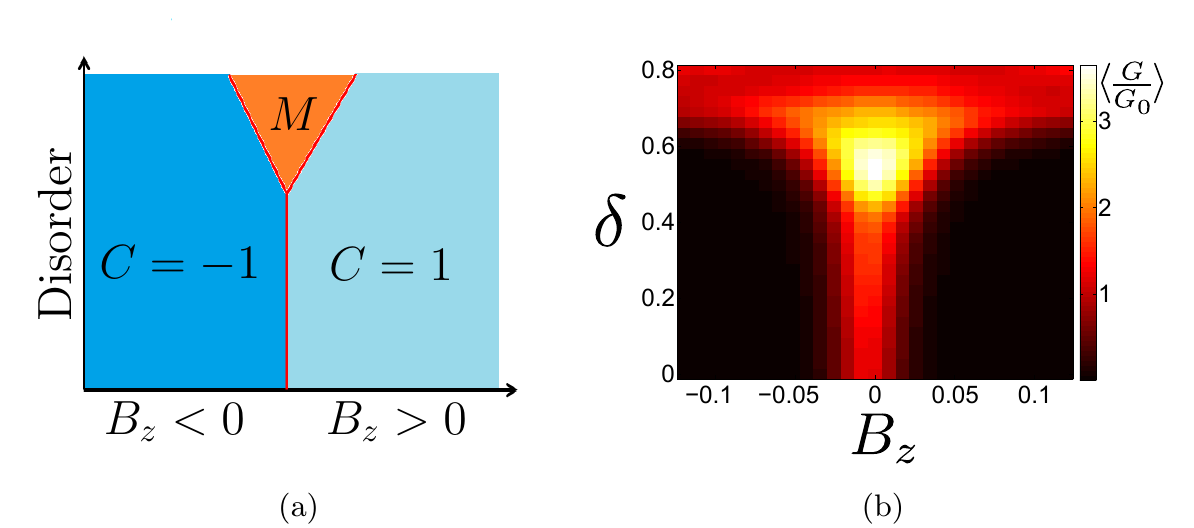}
\caption{ (a) Phase diagram as a function of Zeeman field and disorder strength. Phase transitions are shown in red and $M$ denotes a metallic phase. (b) Disorder averaged bulk thermal conductance of the model in Eq.~\ref{rashba_full}. See \cite{SCparameters} for simulation parameters.\label{fig1}}
\end{figure}

In the phase we analyze, the inclusion of a uniform Zeeman field perpendicular to the 2DEG, $B_z$, leads to a gap opening in the bulk spectrum, and the gapped system possesses a non-trivial Chern number whose sign depends on the sign of $B_z$. Thus, the gapless phase at $B_z=0$ is a transition between two topologically distinct insulating phases. In the presence of disorder, the non-trivial Chern number of the system implies delocalized edge states and localized bulk states. At the transition between different Chern numbers, the bulk gap must close, such that the phase diagram in the space of disorder and gap-opening perturbation contains a critical line, where the bulk states remain delocalized. The topological phase transition occurs at $B_z=0$ also in the presence of weak on-site potential disorder, leading to the phase diagram sketched in \refFig{fig1}a. Starting from the gapless point in the clean limit and increasing disorder strength, the system remains critical; the bulk states remain delocalized while the edge states disappear.

In the following we introduce and study a general scheme that leads to gapless topological superconductors, and then provide a concrete physical realization of such a system. This realization involves magnetically doping the surface of a strong three-dimensional topological insulator (3d TI), and fulfills the necessary requirements without a need for fine tuning.

\textit{\label{GSCmodel}Model ---} 
It was originally shown by Fu and Kane \cite{FuandKane} that when a region of a 3d TI surface is proximity coupled to a ferromagnet and another neighboring region is proximity coupled to an s-wave superconductor, then one-dimensional gapless states must exist at the interface between these two regions. 
Both the ferromagnet and the superconductor induce a gap in the spectrum of the TI surface. The region in space where the gap changes its nature, from a magnetic induced gap to a superconducting induced gap, is the region that hosts the gapless mode. Here, we consider a momentum-space-analogous scenario in which the nature of the gap changes in the two-dimensional Brillouin zone. We show that this construction dictates the existence of gapless excitations in the two-dimensional bulk, which are localized in momentum space and extended in real space.
   
The scheme we consider is based on a family of two-dimensional Hamiltonians of the type,
\begin{align}
 \label{H_model}
H=H_{\rm 0}+H_{\rm m} + H_{\rm SC},
\end{align}
where:
\begin{enumerate}
\item $H_{\rm 0}$ represents a Hamiltonian of spinful electrons, where spin-orbit coupling breaks the degeneracy of the two spin directions for a given momentum. For concreteness, we take a 2DEG with a Rashba spin-orbit coupling 
\begin{align} \label{Rashba_H}
H_{\rm 0}=&\left[ t\left(2-\cos{a k_x}-\cos{a k_y} \right) - \mu \right] \sigma_0
\nonumber
\\
&+ \lambda\left( \sigma_x\sin{a k_x}-\sigma_y\sin{a k_y} \right) ,
\end{align}
where $t$ is the hopping amplitude, $\mu$ is the chemical potential, $\lambda$ is the spin-orbit strength, $a$ is the lattice constant, and the $\sigma_i$'s are the Pauli matrices in spin space. This Hamiltonian has two circular Fermi surfaces, an inner and an outer one.

\item $H_{\rm m}$ is a Zeeman coupling to a spatially periodic magnetization, characterized by a wave-vector $\textbf{Q}$, that opens a gap in parts of the Fermi surface. For concreteness, we take, 
$H_{\rm m}=m\sigma_z\cos{Qx}$, with $Q=2k_F$, where $k_F$ is the Fermi momentum of the outer Fermi surface.

\item $H_{\rm SC}=\Delta\psi_{\mathbf{k},\uparrow} \psi_{-\mathbf{k},\downarrow}+ {\rm H.c.}$ is a superconducting s-wave pairing, with $\psi_{\textbf{k},\sigma}$ being the annihilation operator of a quasi-particle with spin $\sigma$ and momentum $\textbf{k}$, and $\Delta$ is the induced pairing potential.
\end{enumerate}

In the absence of superconductivity, the periodic magnetization, $H_{\rm m}$, defines a new Brillouin zone of size $|\textbf{Q}|$, and opens a gap at some of its edges \footnote{For the practical calculation, we assume that the magnetic periodicity is commensurate with the lattice periodicity}. For $|\textbf{Q}|\sim 2k_F$, where $k_F$ is the Fermi momentum, an open Fermi-surface develops.
The effect of superconducting pairing, $H_{\rm SC}$, depends on its strength. Strong pairing ($|\Delta|>|m|$) renders the entire Fermi surface superconducting, and destroys the effect of the magnetization. Weak superconductivity, on the other hand ($|\Delta|<|m|$), leads to the aforementioned situation where one part of the Fermi surface is gapped by the magnetization and the other part is gapped by superconductivity. 
Therefore, the nature of the gap changes along a path in momentum space that follows the original Fermi surface, and the gap is closed at the point of change.
 
Introducing the Nambu basis, $(\psi_{\uparrow}^{\dagger},\,\psi_{\downarrow}^{\dagger},\,\psi_{\downarrow},-\psi_{\uparrow})$, the full BdG Hamiltonian becomes
\begin{align} \label{rashba_full}
H=& H_{\rm 0} \tau_z + H_{\rm m} \tau_0 + \Delta \sigma_0\tau_x,
\end{align}
where the $\tau_i$'s are Pauli matrices in particle-hole space. The model obeys a particle-hole symmetry ${\cal P}=\sigma_y\tau_y{\cal K}$, ${\cal P}^2=1$, with ${\cal K}$ signifying complex conjugation, and therefore belongs to symmetry class D \cite{class}, which has a topological classification according to $\mathbb{Z}$ in two spatial dimensions.
In the following, we discretize the Hamiltonian \eqref{rashba_full} on a square lattice of $L_x\times L_y$ sites, setting $t=a=1$, $\mu=0$ and $\lambda=\sqrt{2}-1$, which gives $Q=\frac{\pi}{2}$.
\begin{figure}
\centering
\includegraphics[width =\linewidth]{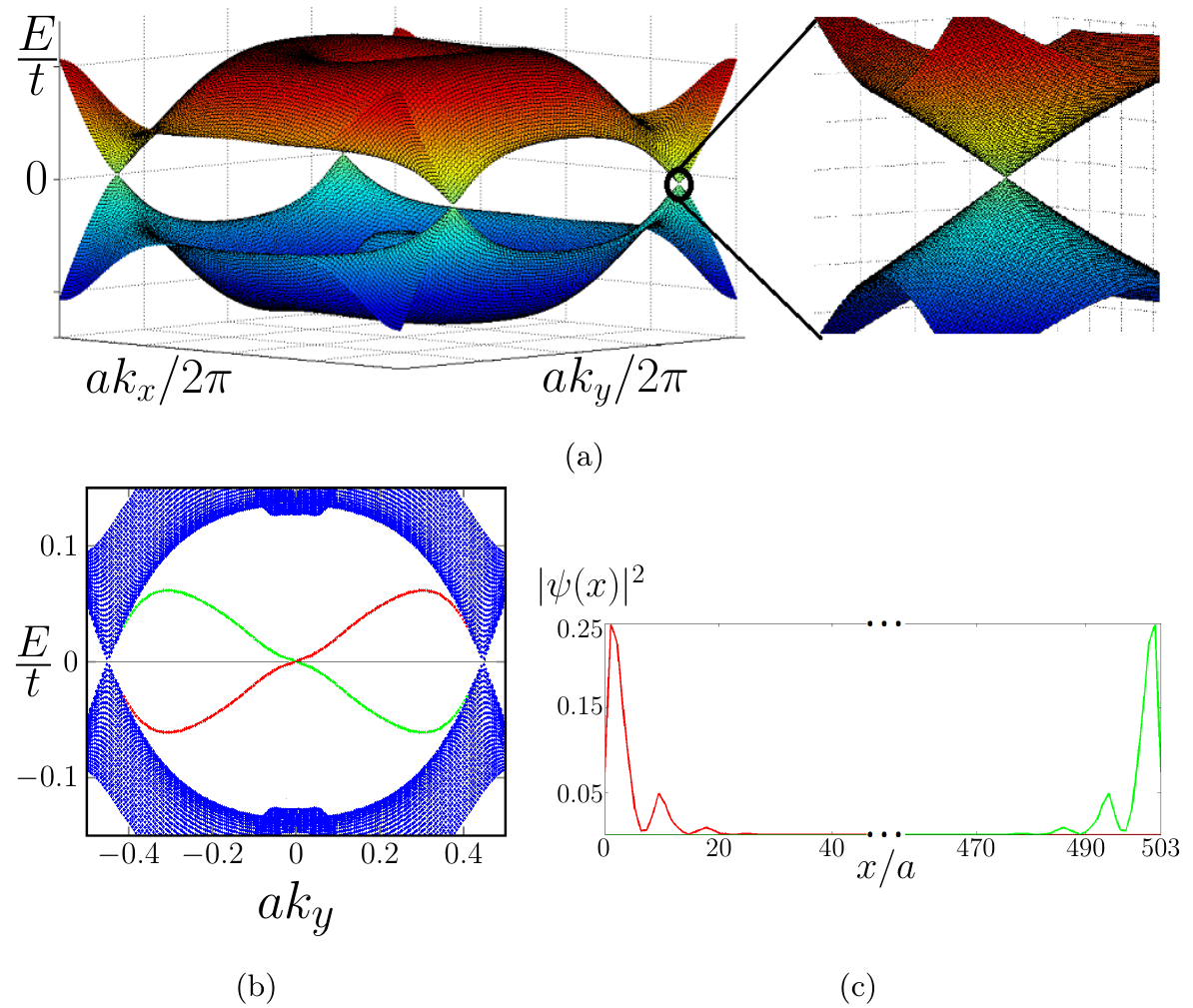}
\caption{ (Color online) Band structure of Hamiltonian \eqref{H_model} with
(a) periodic boundary conditions. (b) An edge along the $y$ direction. Gapless edge states (red, green) coexist with the bulk nodes (blue). 
(c) Typical edge state wave functions.\label{fig2}}
\end{figure}

The band structure near the Fermi level for $|\Delta|<|m|$ is shown in \refFig{fig2}a. There are two distinct Dirac nodes in the spectrum at the edges of the first Brillouin zone. Adding a small uniform Zeeman field, $B_z$, leads to a mass gap at the two Dirac cones. The gapped system is a class $D$ superconductor with a Chern number $C=\mbox{sign}(B_z)$. 
In contrast, for $|\Delta| > |m|$ the spectrum is fully gapped (not shown), with $C=0$ for any $|B_z|<|\Delta|$.
For $B_z=0$ and in the presence of boundaries, we find weak edge states when the boundary is along the $x$-direction and strong edge states for all other orientations of the edge.
The spectrum of the system with boundaries along the $y$-direction and for $\Delta<m$ appears in \refFig{fig2}b. Beside the two bulk nodes at finite $k_y$, there are well localized (strong) chiral edge modes.

\begin{figure}[t!]
\raggedright
{\includegraphics[width=\linewidth]{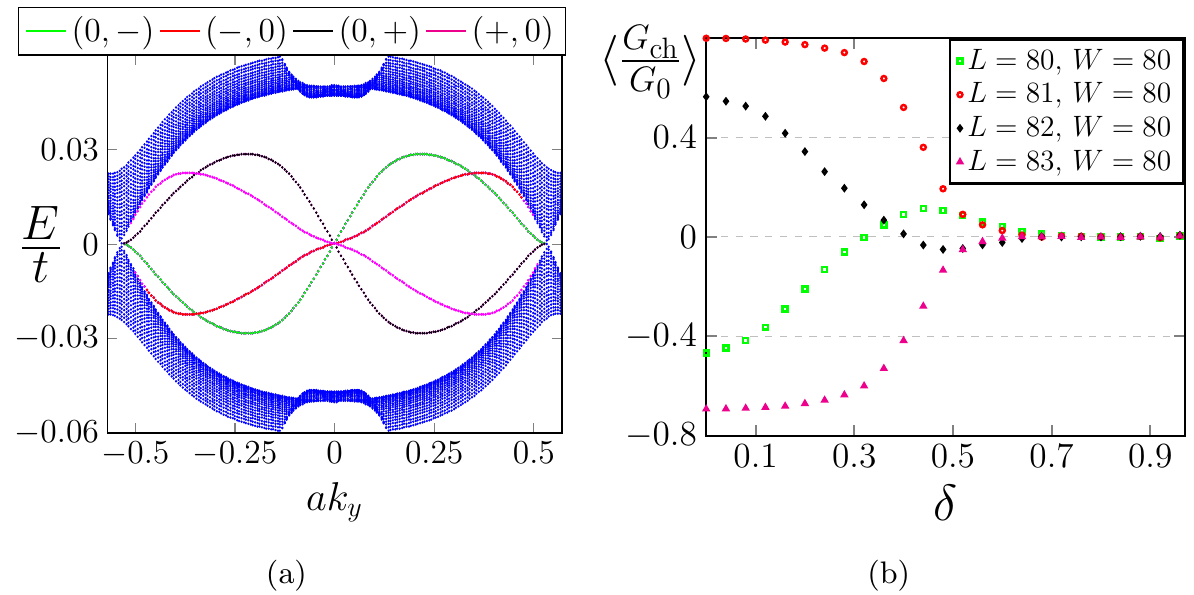}}
\caption{
(Color online) (a) The spectrum of a system with an edge along the $y$ direction for all possible terminations. The colors in the legend denote the right edge state. (b) Strong edge contribution, $G_{\rm ch}=G_{RL}-G_{LR}$, to the thermal conductance as a function of disorder strength $\delta$. We use $m=0.1$ and $\Delta=0.05$. For $\delta=0$, the chirality of the edge states changes depending on the termination and on disorder.\label{fig3}}
\end{figure}

We find that the edge mode properties are not determined uniquely by the edge's orientation. The modulated nature of the magnetization additionally leads to a dependence on the termination of the lattice. For the parameters we chose, the magnetization wave vector is $\textbf{Q}=(\pi/2,0)$, \textit{i.e.}, the magnetization is periodic along the $x$ direction with a periodicity of four sites, and hence, there are four different ways to terminate the lattice in the $x$ direction. We label them according to the magnetization of the last two sites: $(0,+)$, $(+,0)$, $(0,-)$ and $(-,0)$.
For a different choice of magnetic periodicity, the size of the unit cell changes but the physics remains similar. 
The dependence of the edge states on the termination is shown in \refFig{fig3}a, where the spectrum of a system with an edge along the $y$ direction is plotted for the four different terminations. Blue points denote bulk states and other colors denote the right edge state. The left edge state is fixed in the $(0,+)$ termination and is not shown. Both the dispersion and the chirality of the edge state change as the termination changes, but the chirality remains non-zero, at zero energy, as long as particle-hole symmetry is preserved.
Notice that for two terminations the edge states cross the Fermi level three times, once at $k_y=0$ and twice, with an opposite velocity, close to the bulk nodes. Hence, the chiralty is determined by the edge states at the vicinity of the bulk nodes.

We confirm the dependence of the edge modes chirality and their contribution to the thermal conductance on both edge orientation and lattice termination by performing transport simulations. Using a three-terminal geometry, we separate bulk and edge contributions to transport, as described in \refApp{A}A. This can be done both in the clean limit as well as in the presence of disorder, which we model as a spatial variation of the chemical potential in Eq.~\eqref{Rashba_H}, drawn independently for each lattice site from the uniform distribution $[-\delta, \delta]$, with $\delta$ being the disorder strength. 
In the presence of weak disorder, the chiral edge states contribution is slightly reduced due to the hybridization of the edge modes with the bulk.
For moderate disorder, the edge states at finite momenta hybridize strongly with the bulk nodes, as opposed to the edge states at small momenta. Therefore, beyond a certain disorder strength, the edge states around zero momentum dominate the transport, leading to a unique inversion of the edge state chirality (see \refFig{fig3}b) induced by disorder. As the disorder strength is further increased, all edge state contributions go to zero as the system enters a thermal metal phase.

\begin{figure}[t!]
\centering
\includegraphics[width =\linewidth, angle =0] {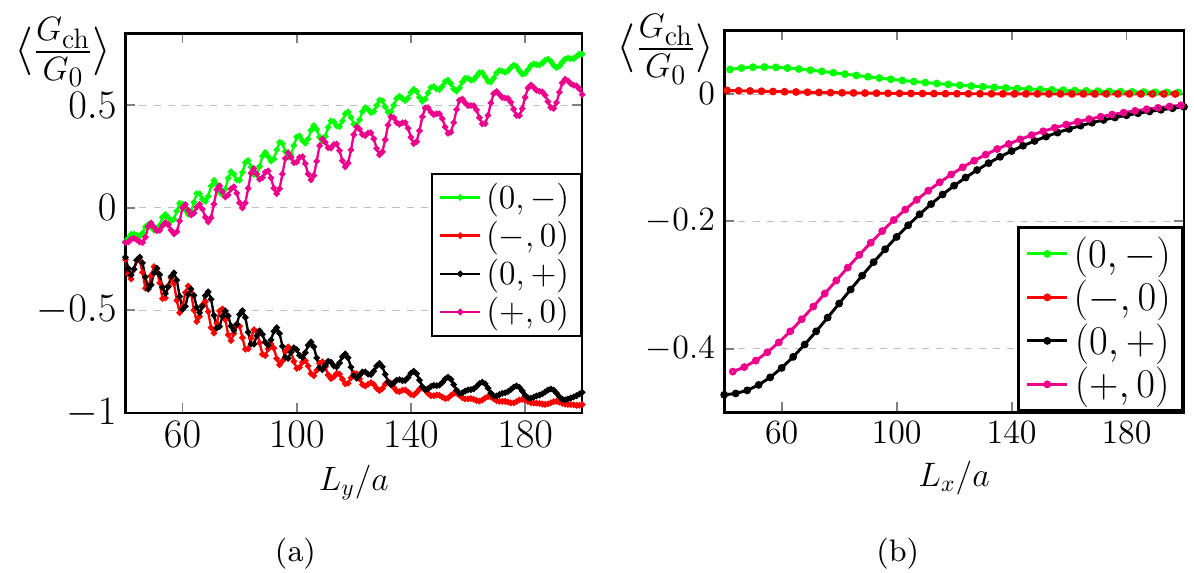}
\caption{Edge contribution, $G_{\rm ch}$, to the thermal conductance in the absence of disorder ($\delta=0$) as a function of system size, for the different terminations. Here $m=0.1$ and $\Delta=0.05$. (a) Strong edge states. While their chirality depends on the termination, their contribution to the conductance approaches unity. (b) Weak edge states. Independently of the termination, their contribution approaches zero.\label{fig4}}
\end{figure}

In the clean case, the contribution of the strong edge states to the thermal conductance becomes quantized in the thermodynamic limit as \refFig{fig4}a shows.
In contrast, the contribution of the weak edge states vanishes as a function of system size, see \refFig{fig4}b. 
In this model, the coupling between the edge and bulk states depends on the normal and superconducting parameters of the model and therefore cannot be independently tuned. The energy scales that correspond to these parameters are large compared to the topological gap scale, $|\Delta|-|m|$, and therefore the weak edge states hybridize effectively.
In the presence of a uniform Zeeman field $B_z$, the dependence on the termination disappears and the chirality is determined solely by the sign of $B_z$. In \refApp{B}B, we provide an illustrative model aiming to intuitively explain the dependance on the lattice termination.  
We plot the bulk thermal conductance of the system as a function of $B_z$ and disorder strength $\delta$ in \refFig{fig1}b. The calculated phase diagram agrees nicely with the theoretical expectation of \refFig{fig1}a. 

\textit{\label{sctnMagneticImpurities}Realization ---}
Class $D$ gapped topological superconductors may be realized in hybrid semiconductor-superconductor structures, by fine-tuning the chemical potential and the applied magnetic field \cite{Rashba1}. We take a different approach, and consider generating complex magnetic order on the surface of a 3d TI without the need for fine tuning.
This may be realized by means of magnetic doping on the surface of the topological insulator 
\cite{LiuLiuXuQiZhang2009MagneticImpuritiesOnTheSurfaceOfATI,SchmidtMiwaFazzio2011SpinTextureAndMagneticAnisotropyOfCoImpuritiesInBi2Se3TI,RosenbergFranz2012SurfaceMagneticOrderingInTopologicalInsulatorsWithBulkMagneticDopants,CapraraTugushevEcheniqueChulkov2012SpinPolarizedStatesOfMatterOnTheSurfaceOfA3DTIWithImplantedMagneticAtoms,ZhuYaoZhangChang2011ElectricallyControllableSurfaceMagnetismOnTheSurfaceOfTopologicalInsulators,YeDingZhaSu2010SpinHelixOfMagneticImpuritiesIn2DHelicalMetal,KlinovajaStanoYazdaniLoss2013TopologicalSuperconductivityAndMajoranaFermionsInRKKYSystems,JiangWu2011SpinSusceptibilityAndHelicalMagneticOrderAtTheEdgesSurfacesOfTopologicalInsulatorsDueToFermiSurfaceNesting,LossRKKYpaper}, which we model as a lattice of magnetic impurities.

\begin{figure}[t!]
\raggedright
{\includegraphics[width=\linewidth]{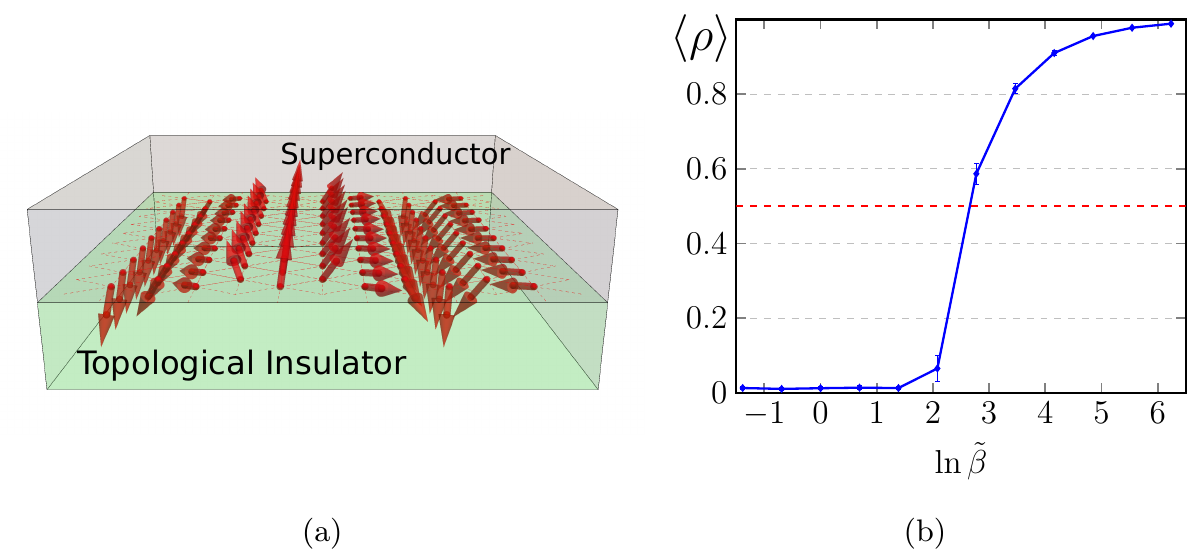}}
\caption{
(Color online) (a) Setup illustration and the spiral spin order in the ground state for an aligned hexagonal lattice of the magnetic impurities. 
(b) Stability of the spiral order $\langle\rho\rangle$ against the normalized inverse temperature $\tilde{\beta}$; both defined in the text.\label{fig5}}
\end{figure}

At low energies, the surface of the undoped system has a Dirac cone, exhibiting spin-momentum locking. As the Fermi level is tuned away from the Dirac point, the Fermi surface becomes nearly hexagonal for a broad range of chemical potentials \cite{Hasan2014,Neupane2012_several_ternary_materials,Miyamoto2012,Sato2010}, a phenomenon called warping. 
Using the model of Fu \cite{Fu2009HexagonalWarpingEffectsInTheSurfaceStatesOfTheTopologicalInsulatorBi2Te3}, the effective
Ruderman-Kittel-Kasuya-Yosida (RKKY) interaction \cite{RudermanKittel1954IndirectExchangeCouplingOfNuclearMagneticMomentsByConductionElectrons,Kasuya1956ATheoryOfMetallicFerroAndAntiferromagnetismOnZenersModel,Yosida1957MagneticPropertiesOfCuMnAlloys,RothZeigerKaplan1966GeneralizationOfTheRKKYInteractionForNonsphericalFermiSurfaces,Nakamura1985_HexagonalFermiSurface_MechanismOf77ReconstructionOnSi111AndGe111SnSurfaces}
between the magnetic impurities is
\begin{align}
\mathcal{H}_{\rm RKKY} = \int_{\rm BZ} d^2q \ S^\lambda_\textbf{q}  J^{\lambda,\lambda^\prime}_\textbf{q} S^{\lambda^\prime}_{-\textbf{q}},
\label{eqRKKYHamiltonianInFourierSpace}
\end{align}
where $\hbar=1$ and $\textbf{S}_\textbf{q}$ is the Fourier transform of the magnetic impurity spin, which we treat as classical.
Here,
\begin{align}
J_\textbf{q}^{\lambda,\lambda^\prime} =\frac{-J_0^2}{2 \pi V_{\rm uc}}\sum_{\textbf{G}}  \chi^{\lambda,\lambda^\prime}_{-(\mathbf{q}+\mathbf{G})}\equiv -\alpha^{-1}\sum_{\textbf{G}}  \chi^{\lambda,\lambda^\prime}_{-(\mathbf{q}+\mathbf{G})},
\label{eqJFourierSpace}
\end{align}
where $J_0$ models an isotropic exchange coupling between the magnetic moments and the surface excitations, $\textbf{G}$ runs over the set of reciprocal lattice vectors of the impurity lattice, and $V_{\rm{uc}}$ is the area of a unit cell.
The quantity $\chi$ is the spin susceptibility of the bare surface without magnetic impurities \cite{BaumStern2012MagneticInstabilityOnTheSurfaceOfTopologicalInsulators,JiangWu2011SpinSusceptibilityAndHelicalMagneticOrderAtTheEdgesSurfacesOfTopologicalInsulatorsDueToFermiSurfaceNesting}, which exhibits pronounced peaks at the six nesting vectors of the hexagonal Fermi surface
 \footnote{ 
The choice of the cutoff in the calculation of $\chi$ has been omitted in previous publications, which we improve upon by a numerical justification in \refApp{C}C.
}.
From \refEq{eqJFourierSpace} we find that when the lattice constant of the magnetic moments exceeds $2\pi |k_F|^{-1}$, the RKKY interaction strongly depends on the impurity lattice structure and on the chemical potential. This allows to engineer the momenta at which the RKKY interaction is peaked, as we exemplify for different lattice structures in \refApp{D}D. For lattice constants smaller than $2\pi |k_F|^{-1}$ however, the RKKY interaction is not significantly altered by the choice of the lattice. 
In this regime, we use a Metropolis algorithm \cite{Metropolis1953} neglecting the contributions to $J$ away from its peaks, and find that the ground state magnetization is a spiral wave, as depicted in \refFig{fig5}a, whose direction and period are determined by one of the nesting vectors. 

To estimate the stability against temperature of this order, we introduce the spiral order parameter $\rho =4\pi M^{-4} \sum_{i=1}^3  S_{Q_i}^2$, where $M^2$ is the number of simulated spins and $Q_i$ are the nesting vectors.
In the spiral ground state $\rho=1$, while it vanishes in the thermodynamic limit for an inverse temperature $\beta \to 0$ \footnote{For a finite number of simulated spins, the lower bound to the order parameter is $\rho = 12 \pi^2/M^4$}.
We calculate the dependence of $\rho$ on $\beta$ for $M=16$ and find a transition to the spirally ordered phase 
at $\tilde{\beta}_c \approx 2.4 \alpha\zeta$, defined as the value at which $\rho=1/2$ (\refFig{fig5}b). Here, $\zeta$ is a material parameter characterizing the warping effect, defined in \refApp{D}D.
To our knowledge, there is no available data for the magnetic couplings $J_0$ on the surface of topological insulators, therefore, we employ a reasonable estimate of $J_0$ ranging from one $\mathrm{meV}$ to about one $\mathrm{eV}$. 
Using these values, the typical range of material parameters \cite{Hasan2014,Neupane2012_several_ternary_materials,Miyamoto2012,Sato2010}, and a spacing of the magnetic impurities of roughly $1 \mathrm{nm}$ - which has been achieved and even underbid in recent experiments \cite{Manoharan2013,Nadj-Perge2014} -, we find a transition temperature between $0.12 K$ and $30 K$. 

After proximity coupling the system to an s-wave superconductor, the phase fulfills the necessary ingredients introduced in the presentation of the general scheme.
For additional details regarding the realization, see Appendices C,D, and E.

\textit{\label{Summary}Summary ---}
We have provided a general scheme for realizing unusual topological superconductors, which simultaneously host gapless Dirac modes in the bulk, and chiral Majorana edge states for almost all edge orientations. 
We analyze the spectral properties and the thermal conductance of the system in the presence of disorder and small Zeeman fields, and find that the structure of the edge states crucially depends on the termination of the lattice.
Specifically, for certain terminations, disorder induces an inversion of the edge state chirality.
Regarding a potential realization, we predict magnetically doped 3d TI in the warping regime to exhibit the proposed phase by magnetic self-organization when proximity coupled to an s-wave superconductor.

\textit{Acknowledgments ---}
Financial support by the DFG (German-Japanese research unit "Topotronics"; priority program SPP 1666 "Topological insulators"), the Helmholtz Foundation (VITI), and the ENB Graduate School on "Topological Insulators" is gratefully acknowledged by TP and BT. TP wants to thank T.~Wehling for correspondence and the Weizmann Institute for hospitality. YB, ICF, and AS thank the European Research Council under the European Union's Seventh Framework Programme (FP7/2007-2013) / ERC Project MUNATOP, the US-Israel Binational Science Foundation and the Minerva Foundation for support.


\section{APPENDIX}

In this appendix, we elaborate on the transport simulations done for the model presented in the main text (Sec.~A) and provide an illustrative model aiming to intuitively explain the edge mode dependence on the lattice termination (Sec.~B).
We give a more detailed discussion on the spiral surface magnetization of strong topological insulators, including the cutoff calculations of the spin susceptibility in Sec.~C, the possibility to engineer the RKKY interaction in Sec.~D, and the stability of the spiral ground state against temperature in Sec.~E.

\subsection{\label{A} A. Transport Simulations}
While both gapped and gapless superconductors are perfect conductors of charge, their thermal conductance is markedly different.
Transport calculations are performed by connecting the system to disorder free leads at temperatures $T_0$ and $T_0+\delta T$, and computing the scattering matrix,
\begin{equation}
 S_{\rm ab}=\begin{pmatrix}
    r_{\rm ab} & t_{\rm ab} \\
    t'_{\rm ab} & r'_{\rm ab}
   \end{pmatrix},
\end{equation}
between any two leads, a and b. This enables us to determine the thermal conductance in the low-temperature, linear response regime, $G_{\rm ab}=G_0{\rm Tr}\,(t^{\phantom\dag}_{\rm ab}t^{\dag}_{\rm ab})$ where $G_0=\pi^2k_{\rm B}^2T_0/6h$ is the quantum of thermal conductance. All transport simulations are performed using the Kwant code \cite{kwant}.

\begin{figure}[t!]
\begin{center}
\includegraphics[width=\linewidth]{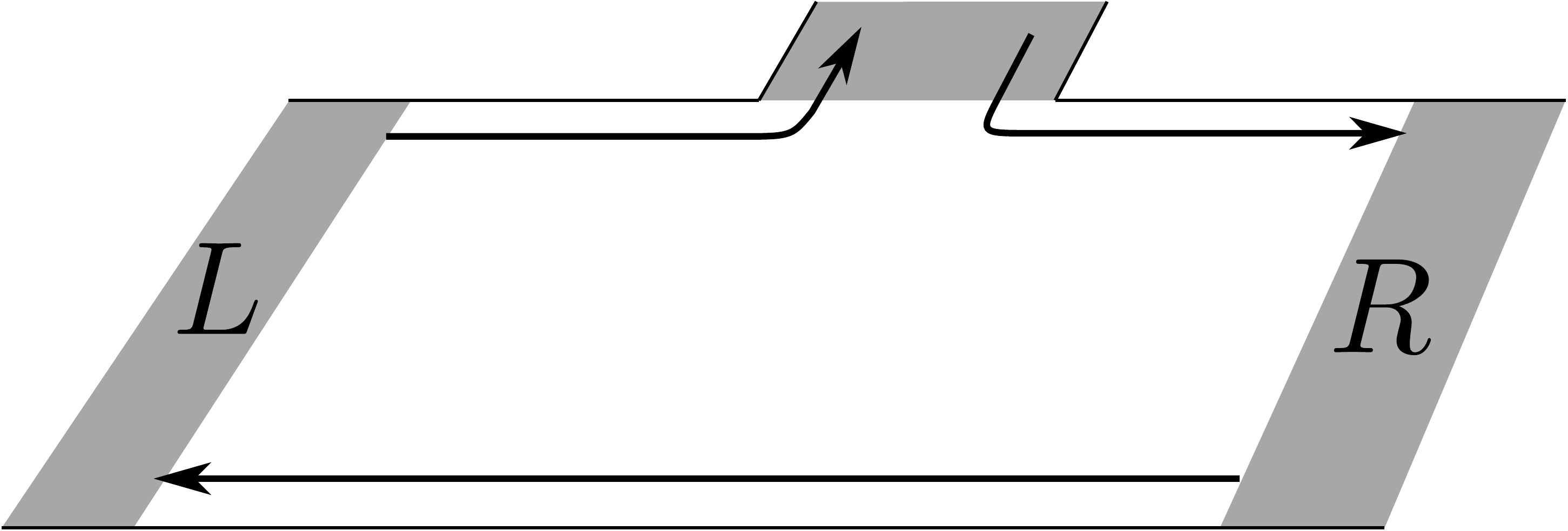}
\caption{ \label{3terminal}%
(Color online) Sketch of the three-terminal geometry used in transport simulations. The edge states (arrows) give the only chiral contribution to transport.}
\end{center} 
\end{figure}

Conventional topological superconductors have a gapped bulk, such that the thermal conductance is only due to edge state transport. In contrast, the model introduced in the main text has both bulk and edge excitations at the Fermi level, so both contribute to the conductance.
In order to separate the bulk and edge contributions, 
we perform transport simulations in a three-terminal setup, as shown in Fig.~\ref{3terminal}. By subtracting the conductance between any two leads from that in the reverse direction ($G_{LR}-G_{RL}$ for instance), we obtain the chiral contribution to transport, which in this model is only due to the edge states.

Transport through the bulk and edge may also be determined by comparing the conductance of a system when changing from hard-wall to periodic boundary conditions, as was done in Ref.~\cite{ourpaper}. In this model however, different lattice terminations independently change the chirality of the edge states, and result in the formation of a spurious conducting channel when periodic boundary conditions are applied. Hence, for a typical termination, the difference in conductance between systems with hard-wall and periodic boundary conditions contains contribution from both the edge and the bulk. Therefore, this setup can not be used for all the possible terminations. The three-terminal setup of Fig.~\ref{3terminal} overcomes this problem, while being less prone to finite-size effects, since boundary conditions are kept fixed.
Additionally, unlike the periodic boundary conditions technique, it has the advantage of modeling an experimentally accessible scenario.

\subsection{\label{B} B. Dependence on Lattice Termination}
\begin{figure}[t]
\begin{center}
   \includegraphics[width =\linewidth] {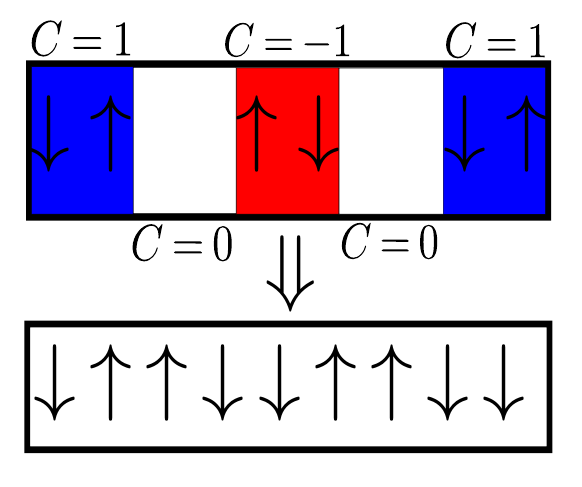}
   \caption{\label{small_Q} Illustration of the connection of our findings to Sau's model \cite{Rashba1} in the limit of small $Q$. Upper part: the system, in real space, is described as stripes of superconductors, with alternating Chern numbers, connected in parallel. The chiral Majorana edges states of each stripe are also depicted. Lower part: the gapped bulk states of each stripe are ignored , and the system may be thought of as a collection of one dimensional chiral channels connected in parallel and coupled by nearest neighbor coupling.}
	 \end{center}
\end{figure}

The dependence of the strong edge states on the magnetic termination may be understood intuitively from the small $|\textbf{Q}|$ limit. It was shown by Sau \emph{et al.}~\cite{Rashba1}, that the combination of a Rashba spin-orbit coupling, s-wave superconducting pairing, $\Delta$, and a uniform Zeeman field, $B_z$, leads to a two-dimensional class $D$ topological superconductor with a well defined Chern number. The value of the Chern number is $\mbox{sign}(B_z)$ for $|\Delta|<|B_z|$ and zero for $|\Delta|>|B_z|$. In the limit of small $|\textbf{Q}|$, the Hamiltonian of Eqs.~(1-3) in the main text can be locally thought of as a system subjected to a uniform Zeeman field. Thus, we can think about the system in real space as stripes of superconductors, with alternating Chern numbers, connected in parallel. Edge states flow where the Chern numbers change. 
Ignoring the gapped bulk states of each stripe, we may view the system as a collection of one dimensional chiral Majorana channels (see an illustration in \refFig{small_Q}).
We assume only a nearest neighbors coupling between these chiral channels, such that channels with the same chirality are coupled by $t_s$ (symmetric coupling) and that channels with an opposite chirality are coupled by $t_a$ (anti-symmetric coupling), and obtain the two dimensional band-structure of the system.
If the symmetric coupling is the dominant coupling, i.e., $t_s>t_a$, each two neighboring Majorana channels with similar chirality form a single Fermionic chiral channel. Coupling Fermionic channels with an alternating chirality through a coupling $t_a$ leads to a gapless phase with two Dirac cones in the two dimensional Brillouin zone.
Then, the gapless bulk states in our original model can be understood as the band structure emerging from  $t_s>t_a$.
The strong edge states are then a remnant of the channels near the edge, and their nature depend periodically on the termination.
Consistently with our findings in the main part of the manuscript, the chirality of the edge states in this picture also depends periodically on the termination and changes sign with a period of two stripes.

Adding a uniform Zeeman field on top of this picture leads to an asymmetry between regions with different Chern numbers. Depending on the sign of the Zeeman field, the size of the regions with a given Chern number increases while size of the regions with the opposite Chern number decreases. In the picture of the one dimensional chiral channels, this can be thought of as dimerization of channels. Hence, a single Chern number prevails in the system.

It should be pointed out that we provided this cartoon model solely as an intuition. Although the model in the main text shows similar features to the cartoon model, it can not be extrapolated from the the small $Q$ picture, since the real space picture in \refFig{small_Q} breaks down as $Q$ becomes comparable to $k_{\rm F}$.      

\subsection{\label{C} C. Calculating the Spin Susceptibility with Proper Cutoffs}

Following Refs.~\cite{BaumStern2012MagneticInstabilityOnTheSurfaceOfTopologicalInsulators,JiangWu2011SpinSusceptibilityAndHelicalMagneticOrderAtTheEdgesSurfacesOfTopologicalInsulatorsDueToFermiSurfaceNesting}, we present a detailed calculation of the surface spin susceptibility in three-dimensional topological insulators. We pay special attention to the choice of the energy cutoff.

\subsubsection{Model}

The two dimensional surface of a three-dimensional topological insulator is well described by the low energy Hamiltonian \cite{Fu2009HexagonalWarpingEffectsInTheSurfaceStatesOfTheTopologicalInsulatorBi2Te3}
\begin{align}
\mathcal{H}_0 = \int d^2 \mathbf{k}\, \mathbf{c}^\dagger_\mathbf{k} \left( v_0 \left( k_x \sigma_y - k_y \sigma_x \right) + \gamma w(\mathbf{k})  \sigma_z \right) \mathbf{c}_\mathbf{k},
\label{eqBareHamiltonian}
\end{align}
where $\sigma_{x,y,z}$ are the Pauli matrices in spin space and $w(\mathbf{k})=(k_+^3+k_-^3)/2$ with $k_{\pm}=k_x \pm \I k_y$. Here, $v_0$ is the electron velocity near the Dirac point, originating from Rashba spin-orbit coupling, and $\gamma$ is the warping
parameter due the cubic Dresselhaus spin-orbit coupling of the bulk.
We choose the basis 
$\mathbf{c}_\mathbf{k} =
\left(
c_{\mathbf{k},\uparrow},
c_{\mathbf{k},\downarrow}
\right)^T
$
with $c_{\mathbf{k},\sigma}$ a fermionic annihilation operator for excitations with momentum $\textbf{k}$ and spin $\sigma$.

\begin{figure}
\includegraphics[width = 0.8 \linewidth]{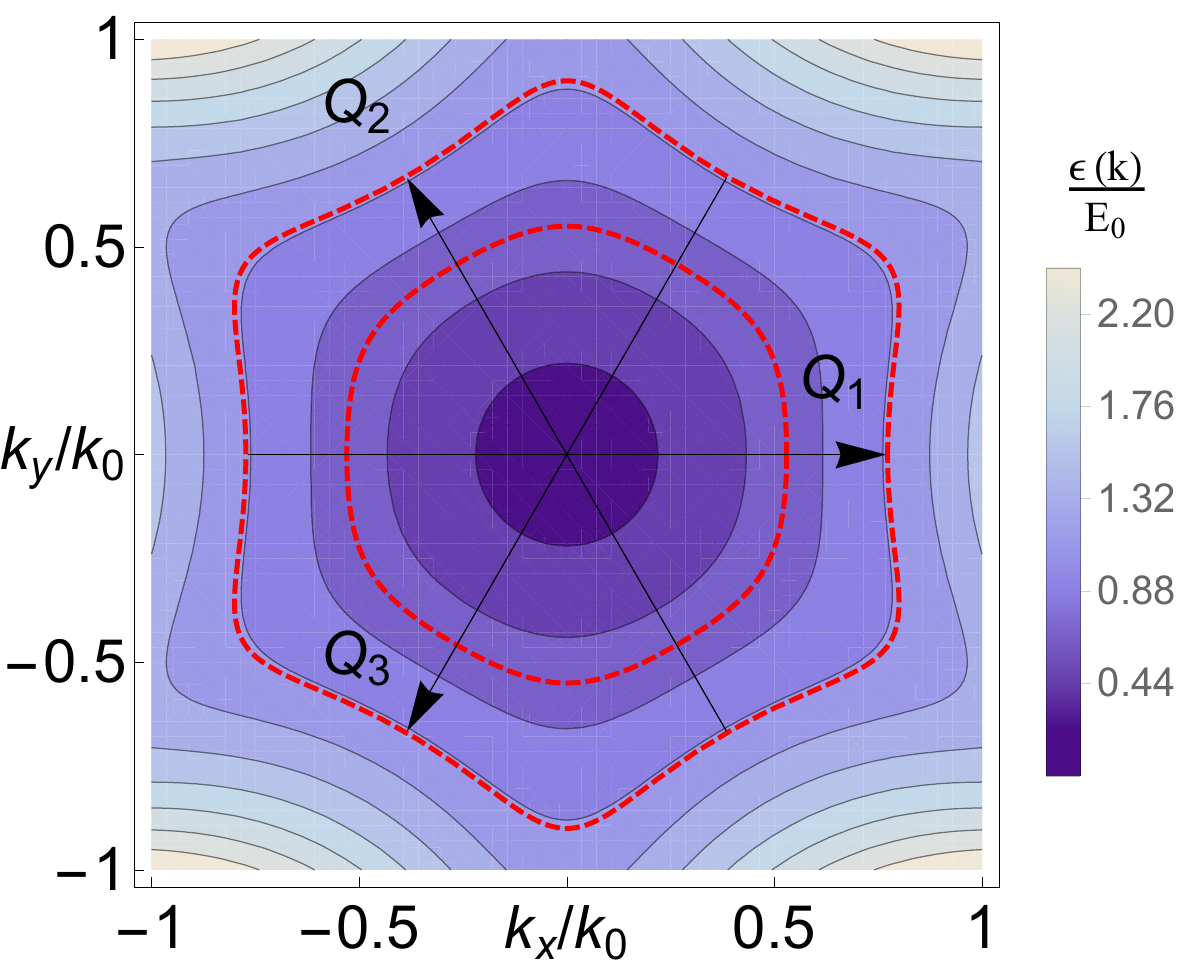}
\caption{
\label{figHexagonalFermiSurfaceRange}(Color online) Fermi surfaces of $\mathcal{H}_0$ in \refEq{eqBareHamiltonian} for various chemical potentials. The initially circular Fermi surface becomes hexagonal with the nesting vectors $\pm(Q_i)_{i=1}^3$. The hexagonal regime is marked by red lines.}
\end{figure}

In the following, we set the unit of energy to $E_0=|v^3_0\gamma^{-1}|^{1/2}$, the unit of momentum to $k_0=|v_0\gamma^{-1}|^{1/2}$ and $\hbar =1$.
Also, for simplicity, we express all energy scales relative to the energy of the Dirac point.
As a function of chemical potential, the circular Fermi surface near the Dirac point becomes hexagonal, and then later develops a 'snowflake' shape. This phenomenon is called warping 
\cite{Fu2009HexagonalWarpingEffectsInTheSurfaceStatesOfTheTopologicalInsulatorBi2Te3}, as shown in \refFig{figHexagonalFermiSurfaceRange}.
For chemical potentials between $0.55 E_0$ and $0.9 E_0$, marked by red contours in \refFig{figHexagonalFermiSurfaceRange}, the Fermi surface is almost hexagonal \cite{BaumStern2012MagneticInstabilityOnTheSurfaceOfTopologicalInsulators}.

In the following, we place a lattice of magnetic impurities on top of the surface, which may be achieved experimentally by means of atomic force microscopy. 
Each lattice position is numbered by two integer indices, $(j_1,\,j_2)=\textbf{j}\in\mathbb{Z}^2$. The spin of the impurity at $\textbf{r}_\textbf{j}$ is described by the operator $S^\lambda_\textbf{j}$.
Magnetic moments couple to the spin density of the itinerant excitations, $s^\lambda_\textbf{j}=c_\textbf{j}^\dagger \sigma^\lambda c_\textbf{j}$, by a local exchange interaction at position 
$\textbf{j}$,
\begin{align}
\mathcal{H}_{int} = \sum_{j \in \mathbb{Z}^2} \sum_{\lambda \in \{x,y,z\}} J^\lambda_j S^\lambda_j s^\lambda_j,
\end{align}
where $J^\lambda_j$ are the exchange coupling constants.
For simplicity, we assume that all moments couple equally to the spin density, and that $J$ is spatially homogeneous: $J^\lambda_j \equiv J_0$.

\subsubsection{RKKY interaction}

If the timescales associated to electron dynamics are considerably shorter than those of the impurities,
the electron spin-density operator can be approximately related to the instantaneous spin configuration of the impurities via linear response theory. Then the system is well approximated by an RKKY Hamiltonian of the form
\begin{align}
\mathcal{H}_{RKKY} = \sum_{i,j} \sum_{\lambda, \lambda^\prime \in \{x,y,z\}} J_0^2 \chi^{\lambda,\lambda^\prime}_{r_i,r_j} S^{\lambda}_i S^{\lambda^\prime}_j.
\end{align}
Here, the non-interacting susceptibility $\chi$ may be expressed in terms of the eigenenergies ($\pm \epsilon_{\textbf{k}}$) and eigenstates ($\psi_{\pm,\textbf{k}}$) of Eq.~(\ref{eqBareHamiltonian}),
\begin{align}
\chi_{r,r^\prime}^{\lambda,\lambda^\prime} =&
				\int_{-\infty}^{\infty} d^2 \textbf{q} \ \frac{e^{- \I \textbf{q}\cdot (\textbf{r} - \textbf{r}^\prime)}}{(2 \pi)^2} \chi^{\lambda,\lambda^\prime}_\textbf{q},
\label{eqSpinSusceptibilityIntegralForm}
\\
\chi^{\lambda,\lambda^\prime}_\textbf{q} =& \int_{-\infty}^{\infty} d^2 \textbf{k}  \sum_{\tau,\rho = \pm} \frac{f_{\beta,\mu}(\tau \epsilon_{k}) - f_{\beta,\mu}(\rho \epsilon_{\textbf{k}+\textbf{q}})}{\tau \epsilon_\textbf{k} - \rho \epsilon_{\textbf{k}+\textbf{q}}}
\nonumber
\\ 
& \times Y^{\lambda,\lambda^\prime}_{\tau,\rho}(\textbf{k},\textbf{k}+\textbf{q})
\theta^{\tau,\rho}_{\textbf{k},\textbf{q},\Lambda_-,\Lambda_+}
\label{eqSpinSusceptibilityFourierSpaceIntegralRepresentation}
\end{align}
where $\epsilon_\textbf{k} = \sqrt{w(k)^2+k^2}$, $f_{\beta,\mu}$ is the Fermi-Dirac distribution function at an inverse temperature $\beta$ with chemical potential $\mu$, and
\begin{align}
Y^{\lambda,\lambda^\prime}_{\tau,\rho}(k,k^\prime) = (\psi_{\tau,\textbf{k}}^{\dagger}\sigma^{\lambda}\psi_{\rho,\textbf{k}'})(\psi_{\rho,\textbf{k}}^{\dagger}\sigma^{\lambda '}\psi_{\tau,\textbf{k}'})^*.
\label{eqConsituentOfSpinSusceptibility}
\end{align}

We define
\begin{align}
\theta^{\tau,\rho}_{k,q,\Lambda_-,\Lambda_+} =
\begin{cases}
1\ \text{if $ \tau \epsilon(k), \rho \epsilon(k+q) \in (\Lambda_-,\Lambda_+)$}
\\
0\ \text{otherwise}
\end{cases}
\end{align}
in terms of energy cutoffs $\Lambda_\pm$, ensuring the validity of the low energy Hamiltonian \eqref{eqBareHamiltonian}. 

\subsubsection{Evaluating the spin susceptibility}\label{appendixNeglectOfPMTerms}
The integral in \refEq{eqSpinSusceptibilityFourierSpaceIntegralRepresentation} contains a sum over four terms, which we label according to the summation indices $\tau$  and $\rho$ as $++$, $\pm$, $\mp$, and $--$.
The $++$ contribution originates from properties close to the Fermi energy; the  $\pm$ and $\mp$ contributions describe high energy processes between the upper and lower branches of the Dirac cone; and the $--$ contribution
describes processes of states that lie deep in the Fermi sea. As such, the latter does not play an important role at low temperatures. Additionally, since bulk states are not close to the Fermi energy nor are they localized at the surface, we do not expect them to contribute significantly to the RKKY interactions.

\begin{figure}
\begin{center}
  \includegraphics[width = 0.8 \linewidth]{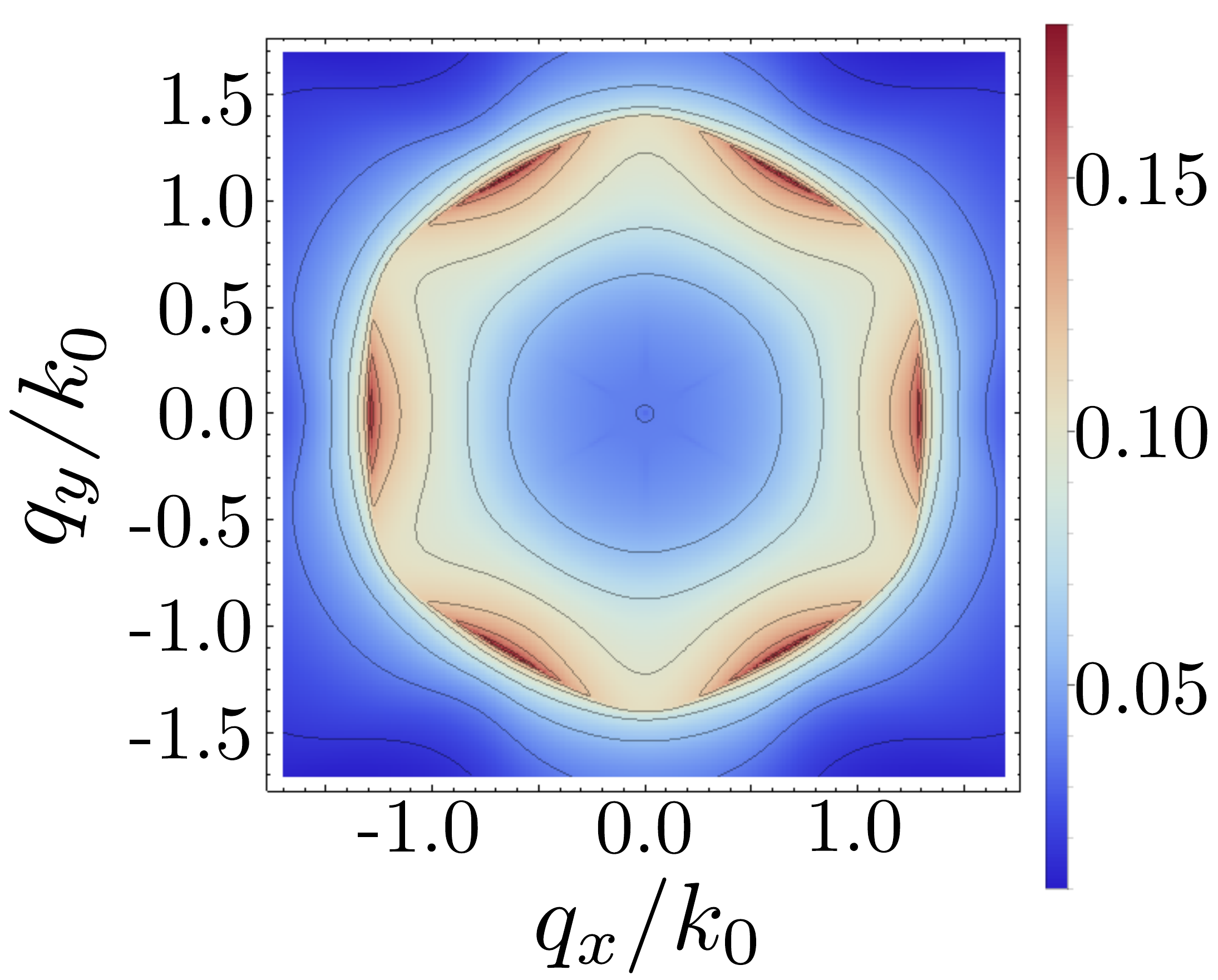}
 \end{center}
\caption{\label{figSpinSusceptibilityPPContribution}(Color online) The largest eigenvalue of the spin susceptibility $\chi$ in momentum space at zero temperature and for $\mu = 0.7$.}
\end{figure}

If the RKKY interaction is dominated by the properties of the system close to the Fermi energy, the $\pm$ and $\mp$ terms may be neglected, as done in Refs.~\cite{BaumStern2012MagneticInstabilityOnTheSurfaceOfTopologicalInsulators,JiangWu2011SpinSusceptibilityAndHelicalMagneticOrderAtTheEdgesSurfacesOfTopologicalInsulatorsDueToFermiSurfaceNesting}.
This leads to the spin susceptibility shown in \refFig{figSpinSusceptibilityPPContribution}, where only the $++$ term is considered. The dominant contributions occur close to the six nesting vectors (\refFig{figHexagonalFermiSurfaceRange}).
The data is obtained by numerically integrating \refEq{eqSpinSusceptibilityFourierSpaceIntegralRepresentation} over a $1600 \times 1600$ hexagonal lattice with a resolution of $100 \times 100 /  k_0^2$. The integrand of \refEq{eqSpinSusceptibilityIntegralForm} is made well-defined at the boundaries of the integration region by regularizing the denominator with $\eta = 2^{-13} E_0$.

\begin{figure}
\begin{center}
\begin{subfigure}{0.3 \linewidth}
\includegraphics[width= \textwidth]{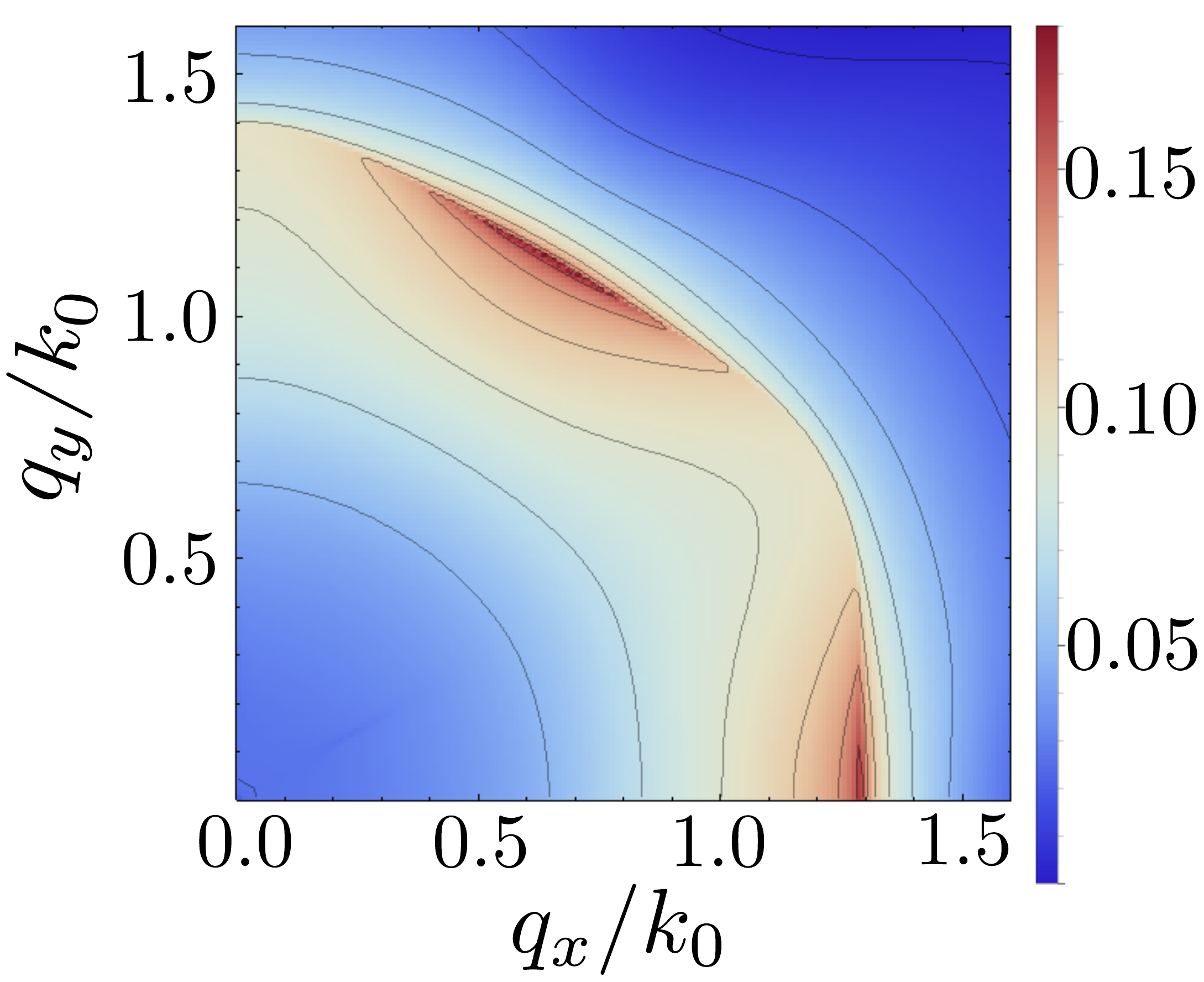}
\caption{$\Lambda_+ = - \Lambda_- = 0$}
\end{subfigure}
\begin{subfigure}{0.3 \linewidth}
\includegraphics[width= \textwidth]{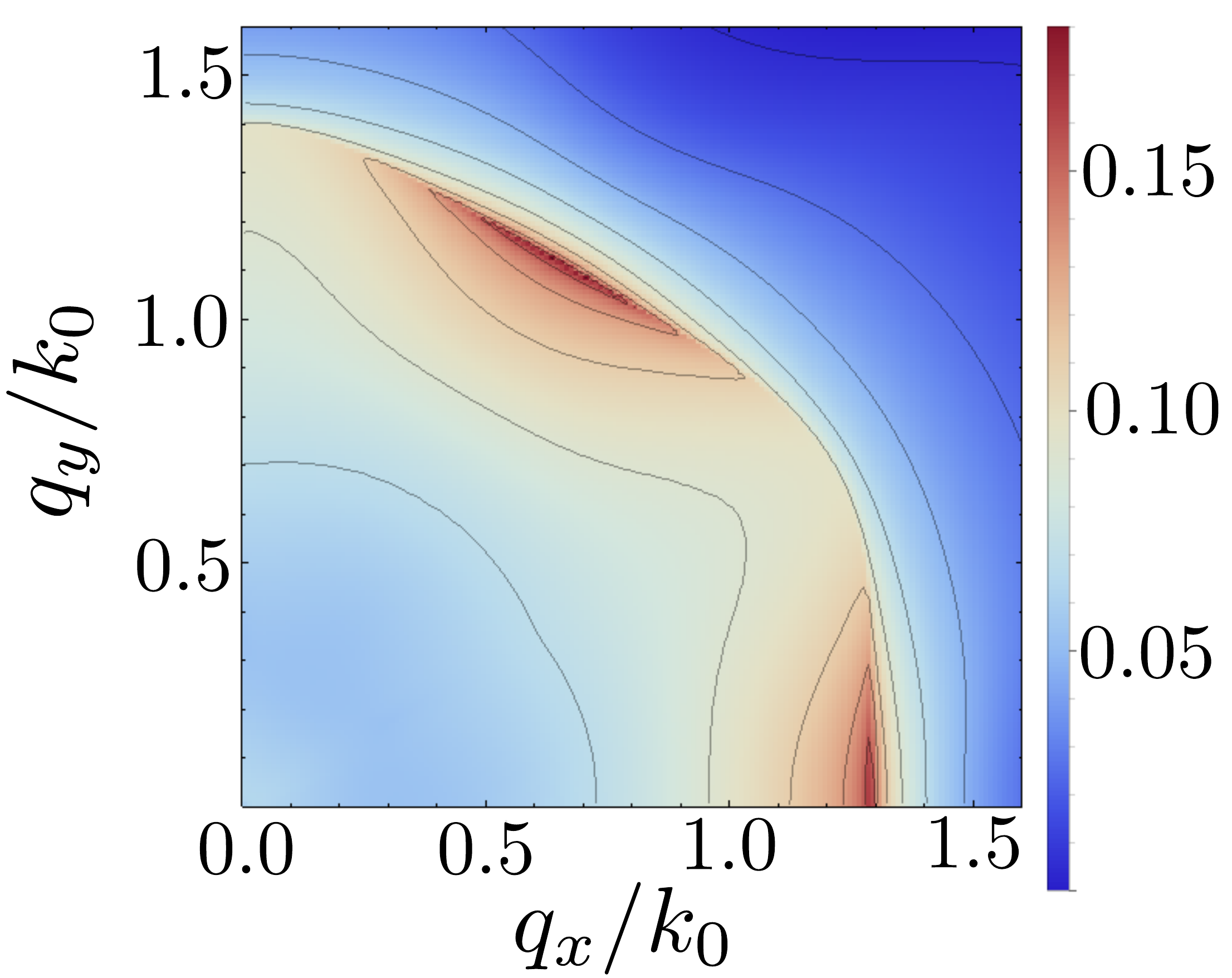}
\caption{$\Lambda_+ = - \Lambda_- = 1 E_0$}
\end{subfigure}
\begin{subfigure}{0.3 \linewidth}
\includegraphics[width= \textwidth]{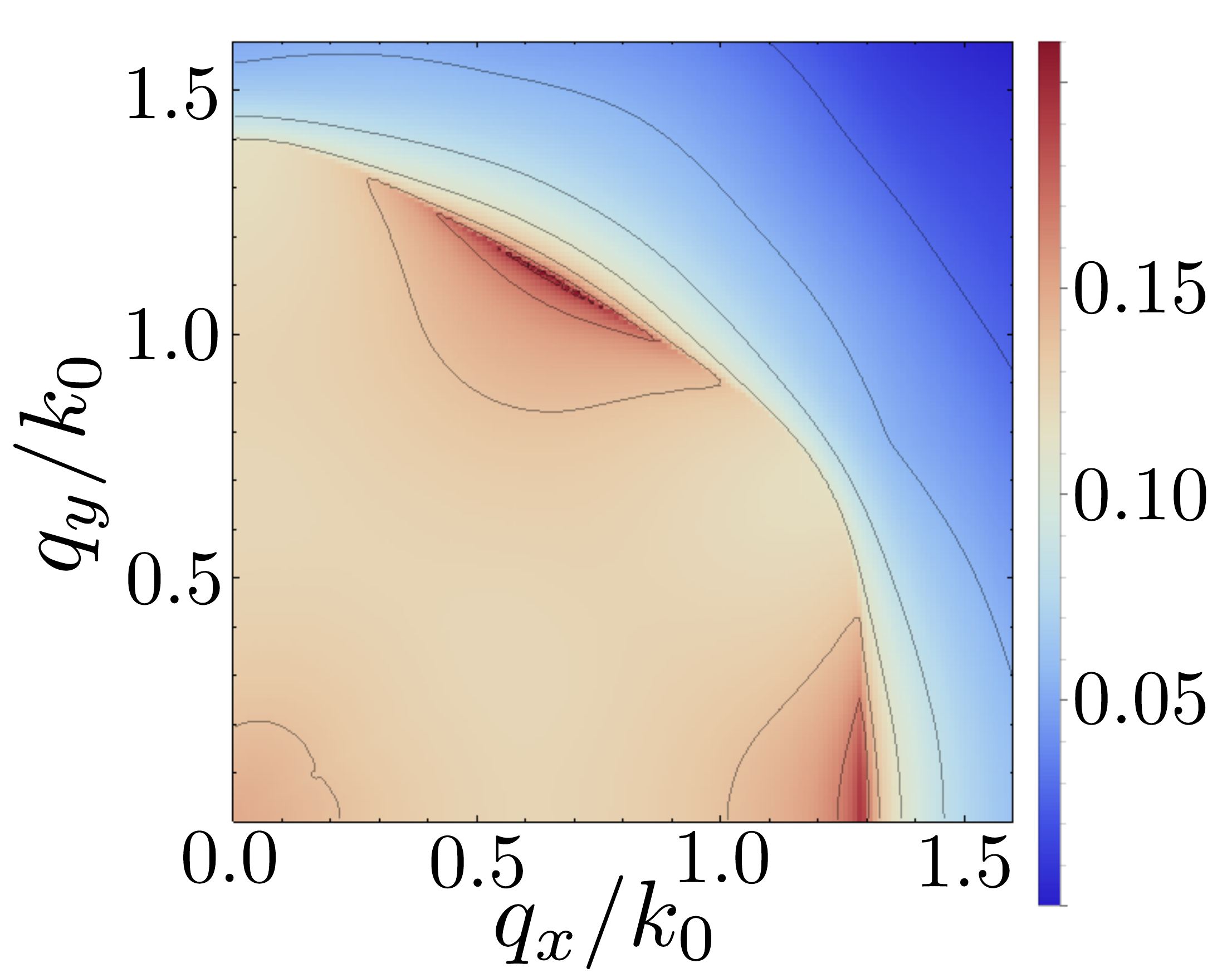}
\caption{$\Lambda_+ = - \Lambda_- = 2 E_0$}
\end{subfigure}
\end{center}
\caption{\label{figChiLargestEV}(Color online) The largest eigenvalue of the spin susceptibility $\chi$ for different cutoffs $\Lambda$ at zero temperature and $\mu = 0.7$. For large cutoffs, the concentration of $\chi$ around the nesting momenta gets reduced and the low momenta contributions have to be taken into account.}
\end{figure}

The mathematical structure of the $\pm/\mp$ contributions, however, does not allow to neglected them a priori. In fact,
without the introduction of the cutoffs $\Lambda_\pm$, the spin susceptibility would be dominated by these contributions.
We give numerical estimates for the range of cutoffs in which their omission is valid.
\refFig{figChiLargestEV} shows the effect of an increasing cutoff, from $0$ to $2E_0$, for $\Lambda = \Lambda_+ = -\Lambda_-$.
The $\pm/\mp$ terms contribute to the susceptibility at small momenta, such that keeping only the contributions close to the nesting vectors is not justified for sufficiently large $\Lambda$.

Recent experimental data \cite{Hasan2014,Neupane2012_several_ternary_materials,Miyamoto2012,Sato2010} shows that for most materials the upper cutoff is larger than the absolute value of the lower one and that both lie within the range where the $\pm$/$\mp$ contributions may be omitted.
In Bi$_2$Te$_2$Se for instance, the lower cutoff is close to zero  \cite{Miyamoto2012,Neupane2012_several_ternary_materials}. In GeBi$_2$Te$_4$ however, both cutoffs are larger than for most materials \cite{Neupane2012_several_ternary_materials}.

\newlength{\smallLength}
\setlength{\smallLength}{0.3 \linewidth}

\subsection{\label{D} D. Engineering Peak Positions in the RKKY Interaction}
The position of the peaks that dominate the RKKY interaction in momentum space can be engineered by choosing different lattice structures, lattice constants, and chemical potentials.
The dependence of the RKKY interaction $J$ on the spin susceptibility (\refEq{eqJFourierSpace} of the main text) reads

\begin{align}
J_q^{\lambda,\lambda^\prime} =\frac{-J_0^2}{2 \pi V_{uc}}\sum_{\textbf{G}}  \chi^{\lambda,\lambda^\prime}_{-(\mathbf{q}+\mathbf{G})},
\label{eqJFourierSpace2}
\end{align}
where $\textbf{G}$ is the set of the reciprocal lattice vectors and $V_{uc}$ is the unit cell area of the magnetic impurity lattice.
Given the spin susceptibility, we consider $J$ for different lattice structures and lattice constants.
For $a < \pi Q^{-1}$, where $Q$ is the modulus of the nesting vectors, the choice of the lattice structure has no significant effect on the RKKY interaction.
However, when $a>\pi Q^{-1}$, different peaks overlap since the nesting vectors exceed the first Brillouin zone. This allows the position of the peaks to be engineered.
Exemplary engineered peak positions are shown in Fig.~\ref{figJForDifferentLattices}.
There, the largest eigenvalue of the RKKY interaction is shown for a a square lattice (top panels), a hexagonal lattice with the lattice vectors $(1,0)a$ and $(1/2,\sqrt{3}/2)a$ (middle panels), and a $\pi/3$ rotated hexagonal lattice (bottom panels).
Changing the lattice constant has a similar effect to tuning the chemical potential within the hexagonal range. The latter procedure has the advantage of leaving the sample unaltered and is more experimentally accessible.

\newcommand{\raiselength}{-0.43}
\newcommand{\graphicwidth}{0.93}
\newcommand{\legendwidth}{0.07}
\newcommand{\legendheight}{0.75}

\begin{figure}

  \begin{subfigure}{\smallLength}
    \raisebox{-0.5 \height}{\includegraphics[width=\graphicwidth \textwidth]{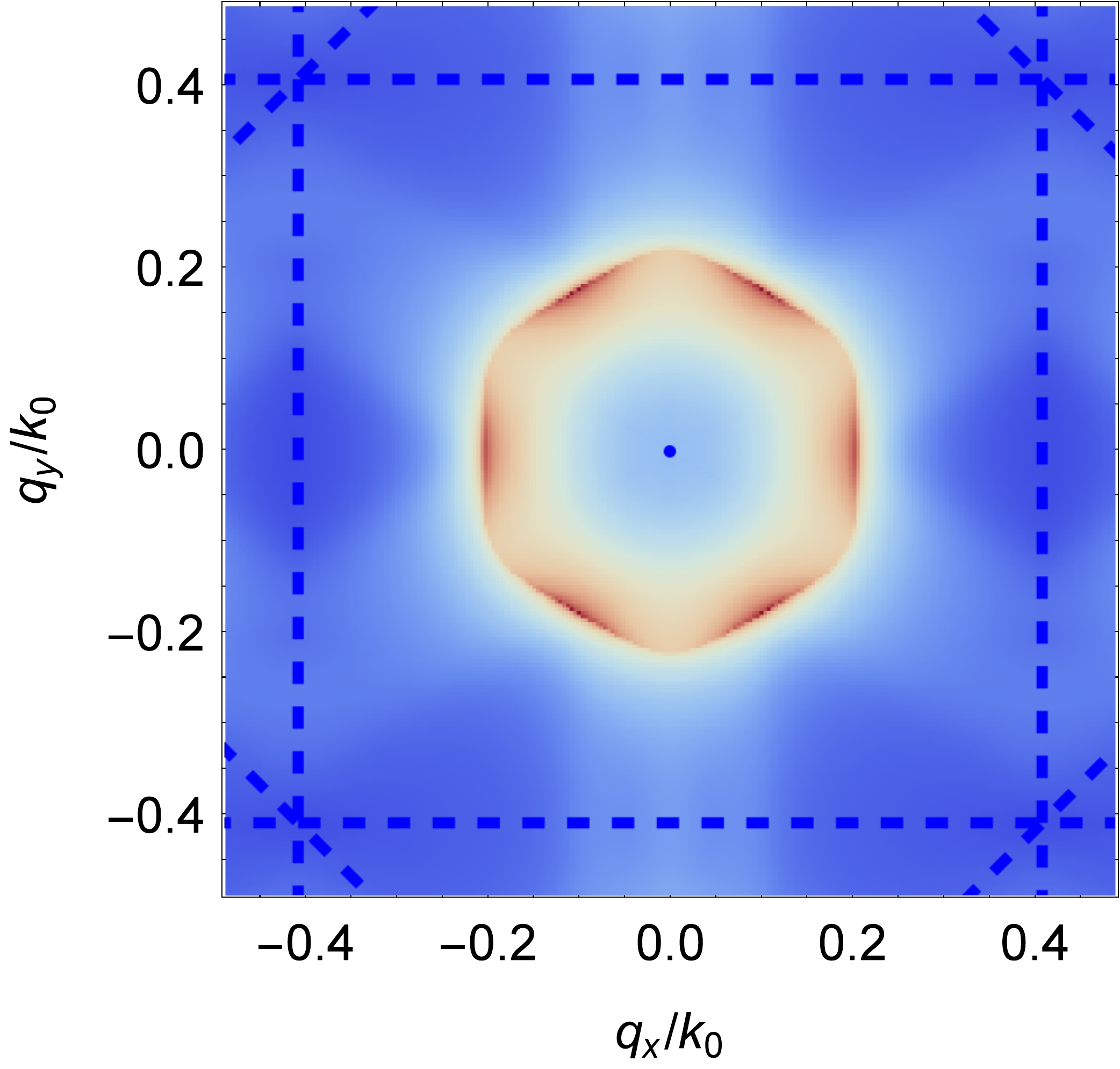}}\raisebox{\raiselength\height}{\includegraphics[height=\legendheight\textwidth]{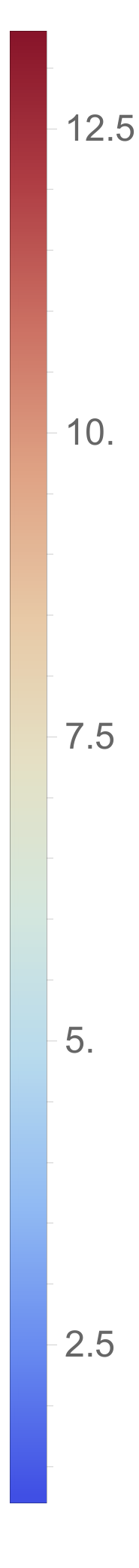}}
    \caption{\label{figJLargestEVLatticeSquare0p25}$a= \frac{\pi}{2}Q^{-1}$}
  \end{subfigure}
  \begin{subfigure}{\smallLength}
    \raisebox{-0.5 \height}{\includegraphics[width=\graphicwidth \textwidth]{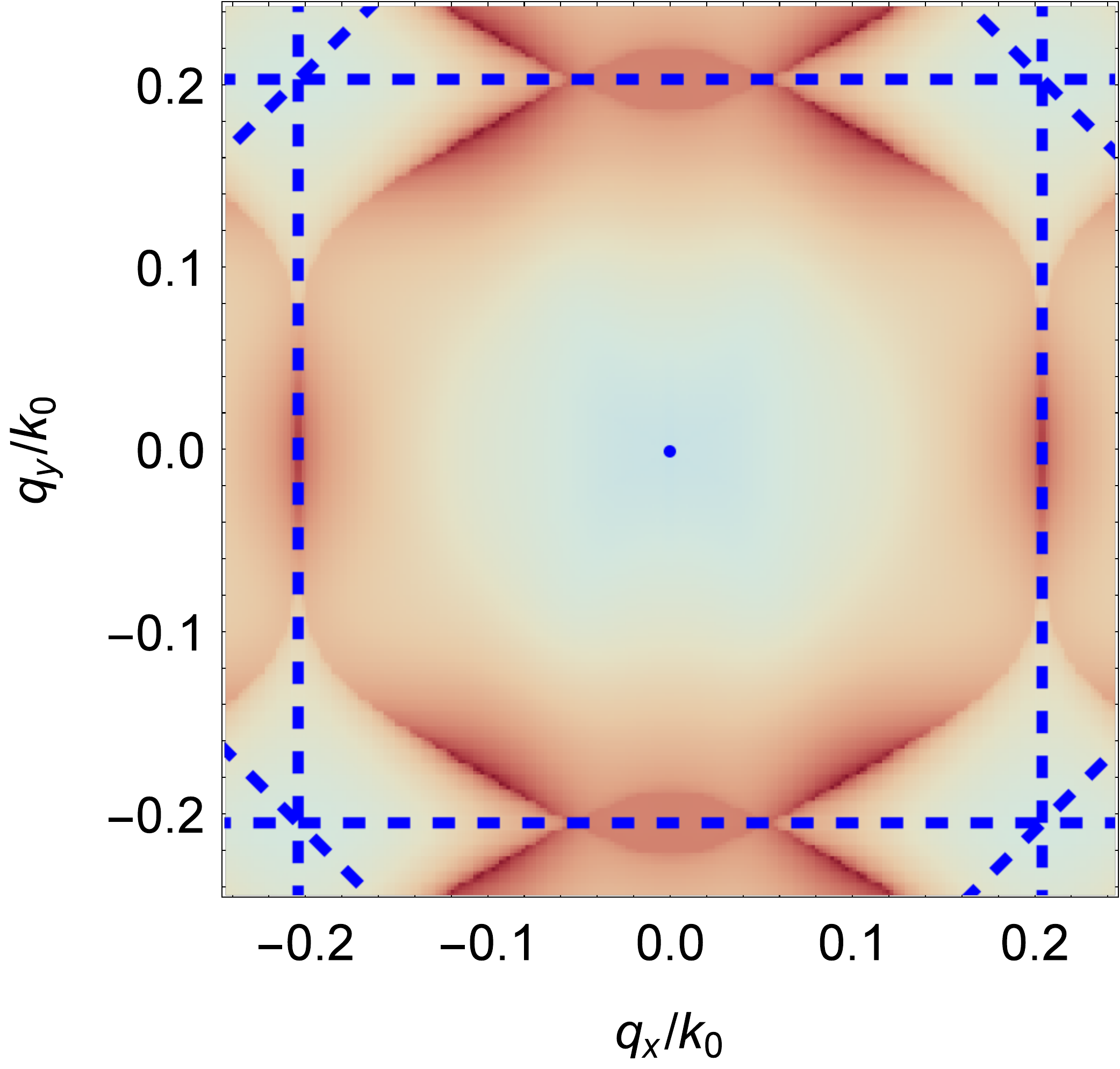}}\raisebox{\raiselength\height}{\includegraphics[height=\legendheight\textwidth]{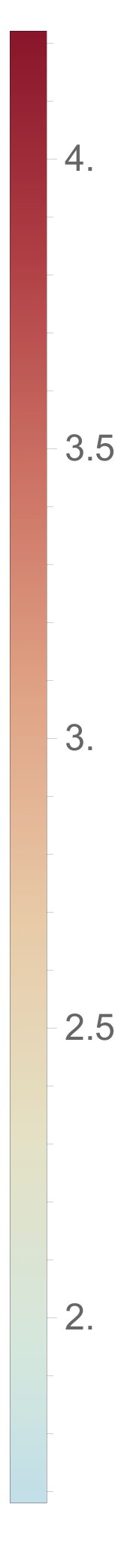}}
    \caption{\label{figJLargestEVLatticeSquare0p5}$a= \pi Q^{-1} $}
  \end{subfigure}
  \begin{subfigure}{\smallLength}
        \raisebox{-0.5 \height}{\includegraphics[width=\graphicwidth \textwidth]{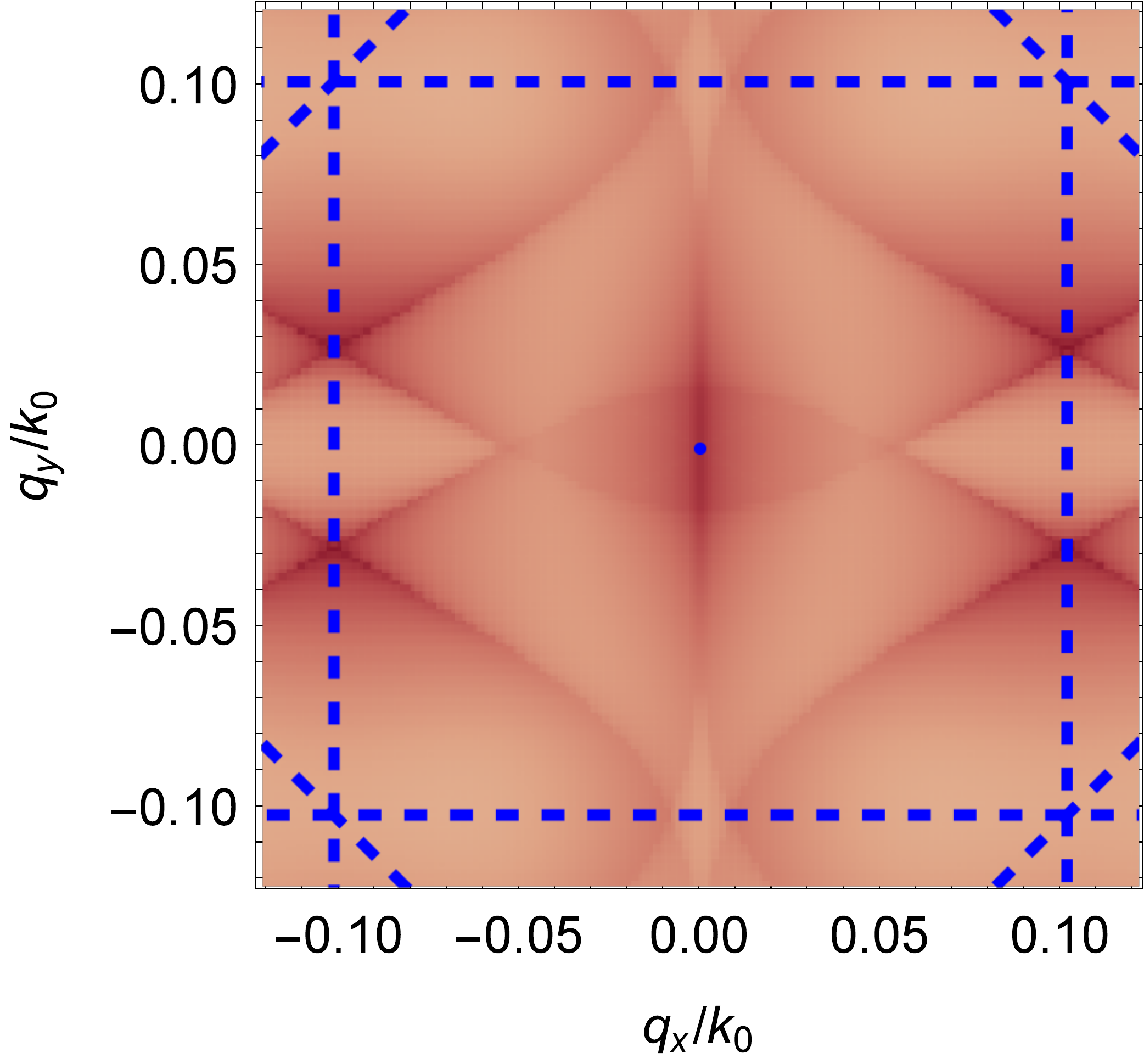}}\raisebox{\raiselength\height}{\includegraphics[height=\legendheight\textwidth]{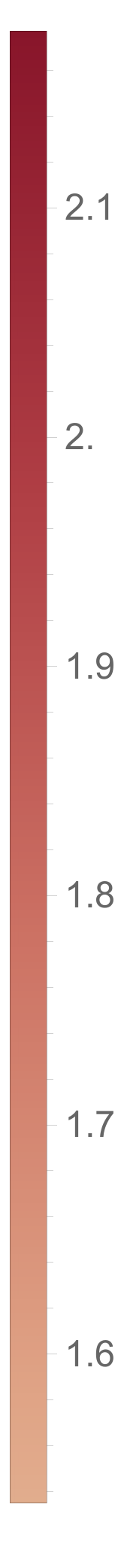}}
    \caption{\label{figJLargestEVLatticeSquare1p}$a= 2 \pi Q^{-1}$}
  \end{subfigure}

  \begin{subfigure}{\smallLength}
        \raisebox{-0.5 \height}{\includegraphics[width=\graphicwidth \textwidth]{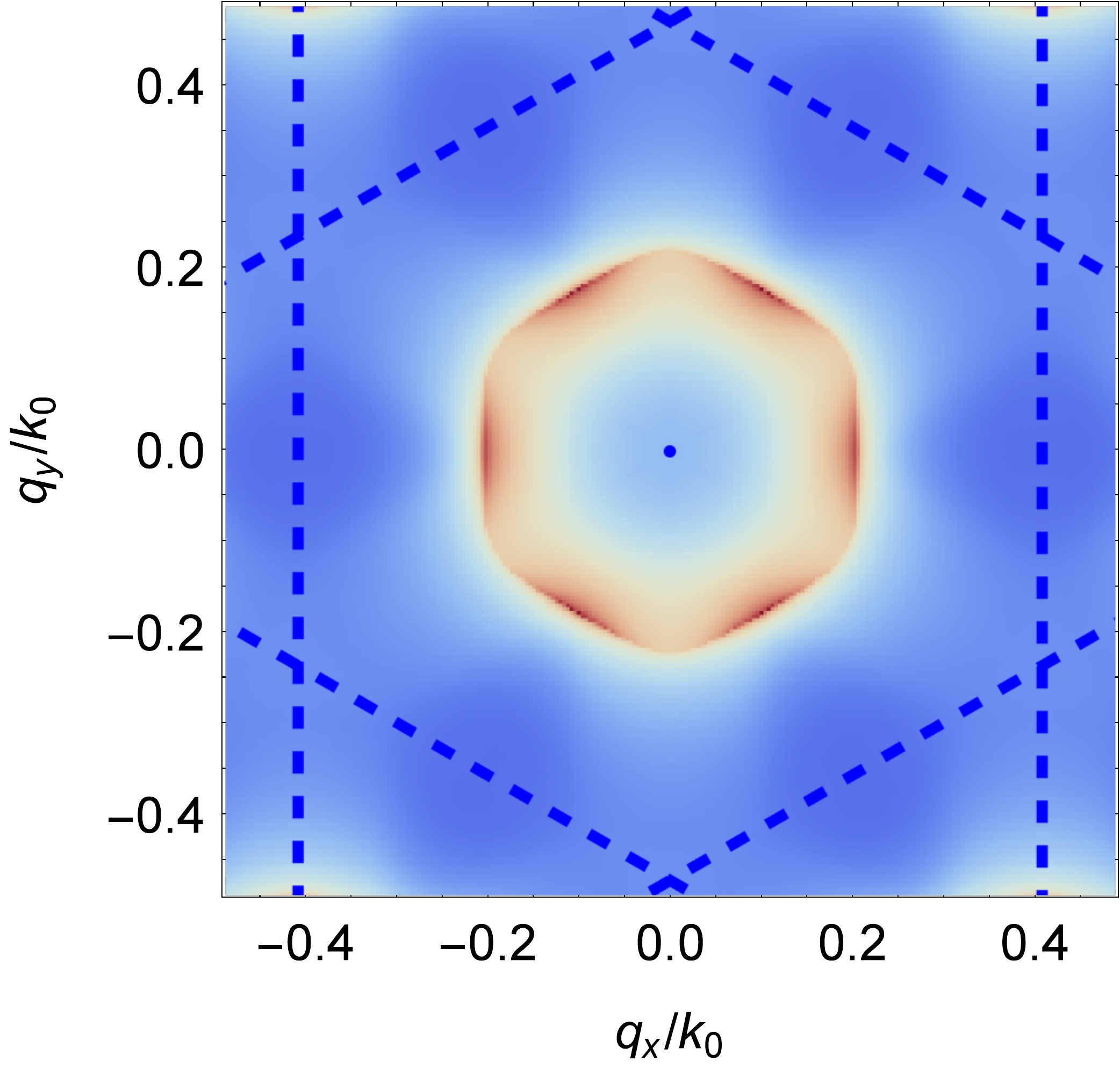}}\raisebox{\raiselength\height}{\includegraphics[height=\legendheight\textwidth]{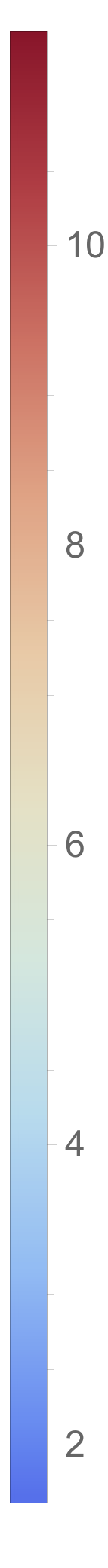}}
    \caption{\label{figJLargestEVLatticeHexAligned0p25}$a= \frac{\pi}{2}Q^{-1}$}
  \end{subfigure}
  \begin{subfigure}{\smallLength}
        \raisebox{-0.5 \height}{\includegraphics[width=\graphicwidth \textwidth]{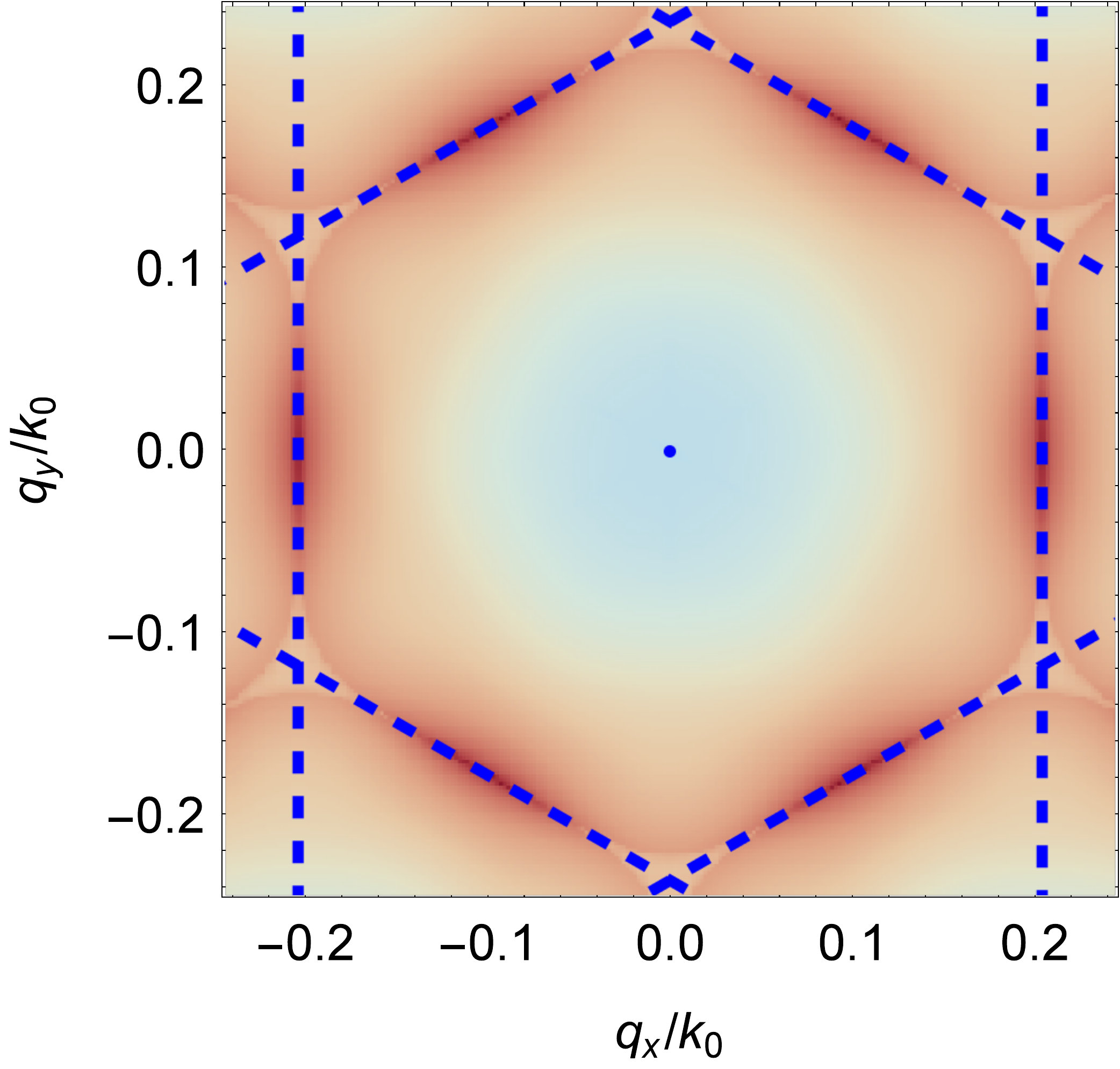}}\raisebox{\raiselength\height}{\includegraphics[height=\legendheight\textwidth]{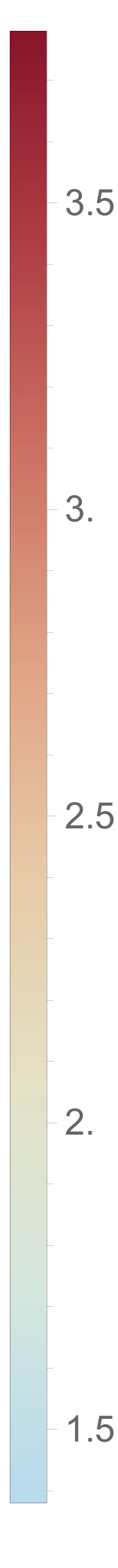}}
    \caption{\label{figJLargestEVLatticeHexAligned0p5}$a= \pi Q^{-1}$}
  \end{subfigure}
  \begin{subfigure}{\smallLength}
        \raisebox{-0.5 \height}{\includegraphics[width=\graphicwidth \textwidth]{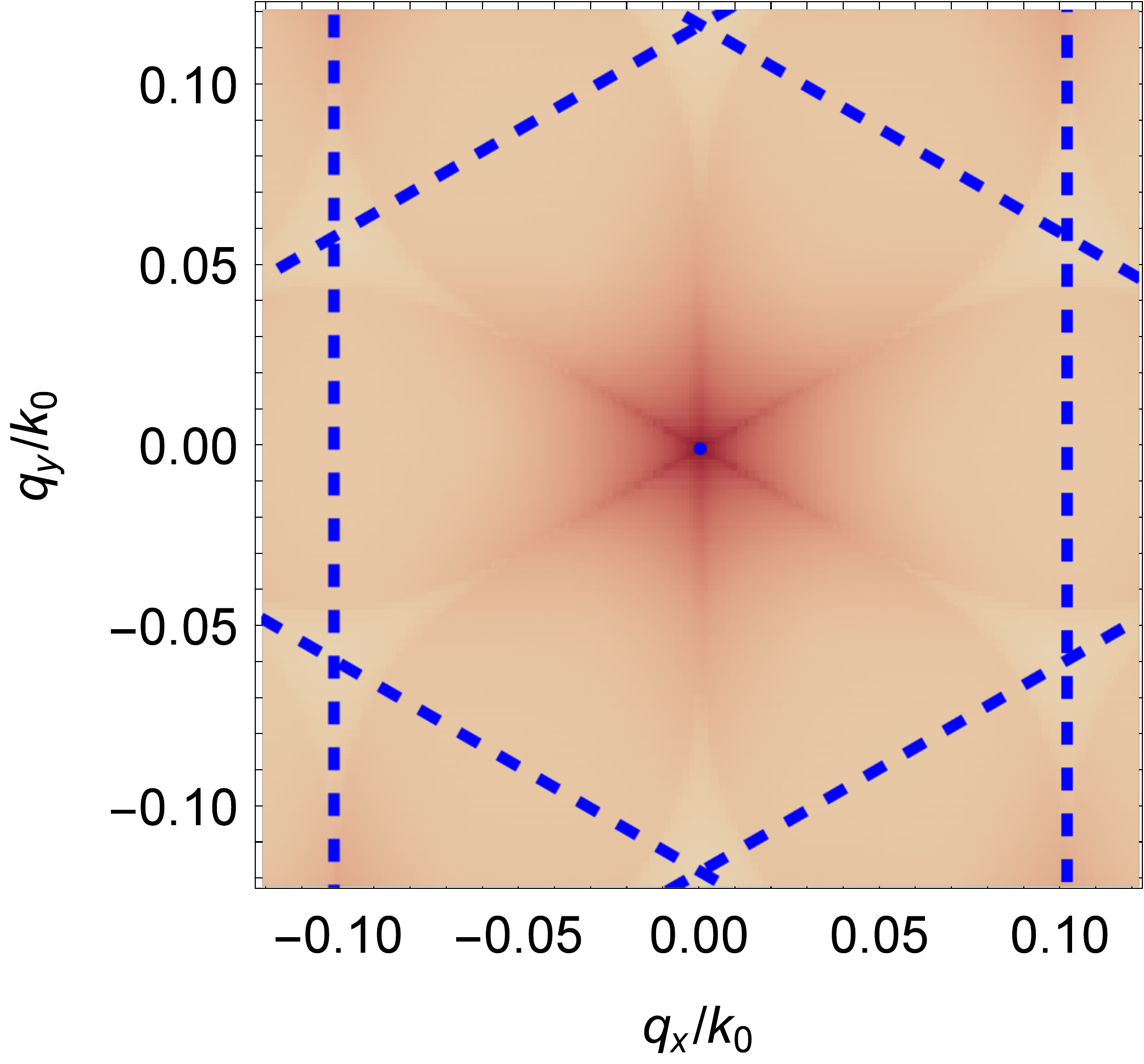}}\raisebox{\raiselength\height}{\includegraphics[height=\legendheight\textwidth]{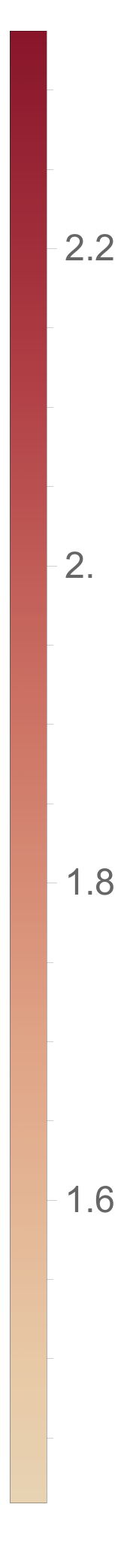}}
    \caption{\label{figJLargestEVLatticeHexAligned1p}$a= 2 \pi Q^{-1}$}
  \end{subfigure}

  \begin{subfigure}{\smallLength}
        \raisebox{-0.5 \height}{\includegraphics[width=\graphicwidth \textwidth]{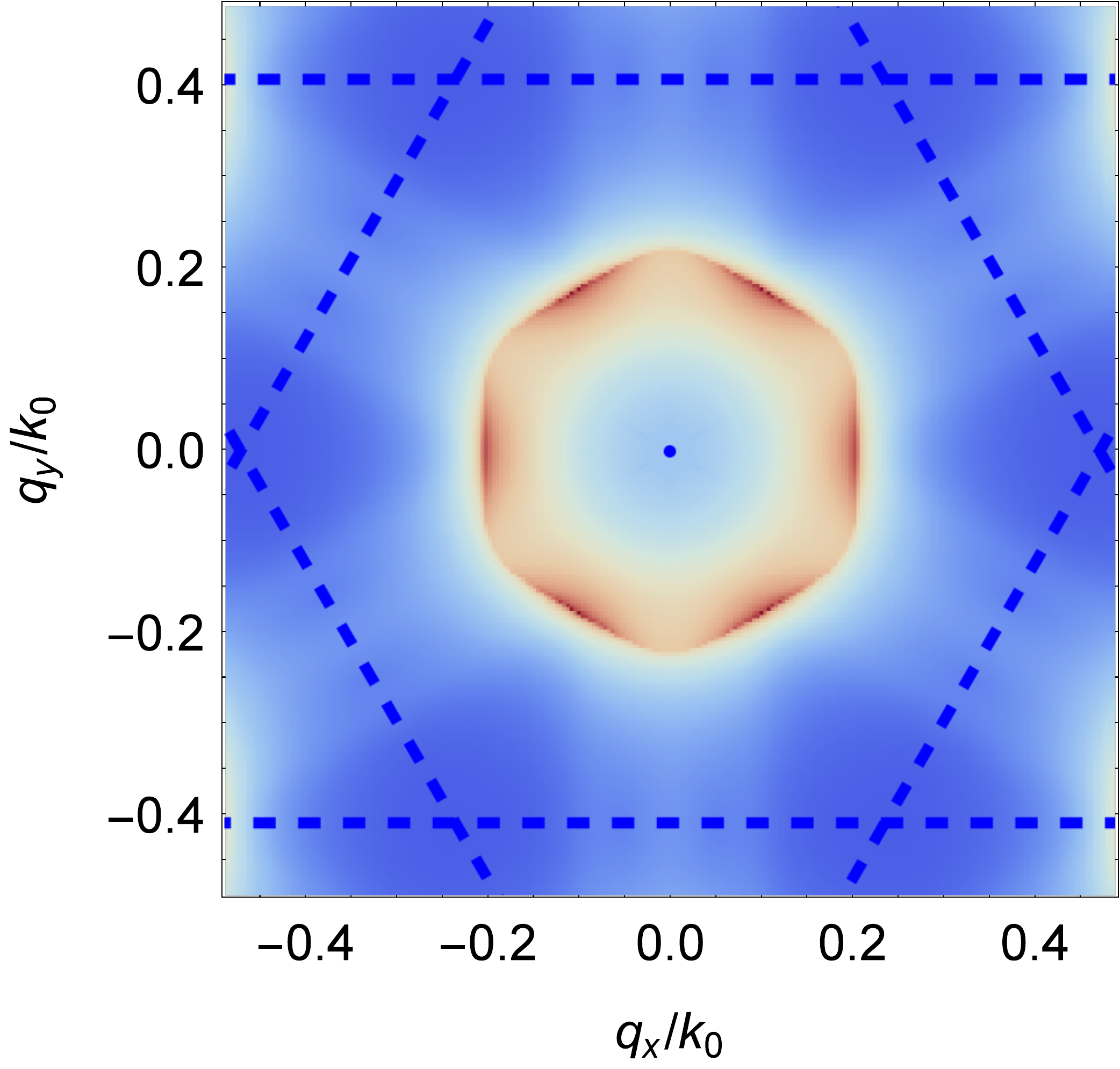}}\raisebox{\raiselength\height}{\includegraphics[height=\legendheight\textwidth]{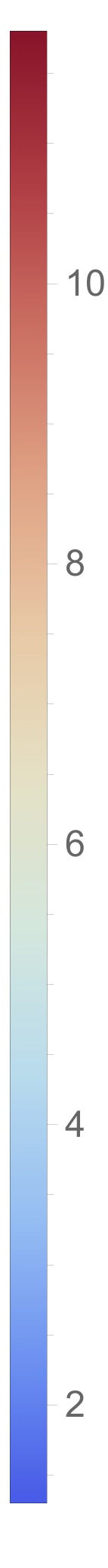}}
    \caption{\label{figJLargestEVLatticeHexContraAligned0p25}$a= \frac{\pi}{2}Q^{-1}$}
  \end{subfigure}
  \begin{subfigure}{\smallLength}
    \raisebox{-0.5 \height}{\includegraphics[width=\graphicwidth \textwidth]{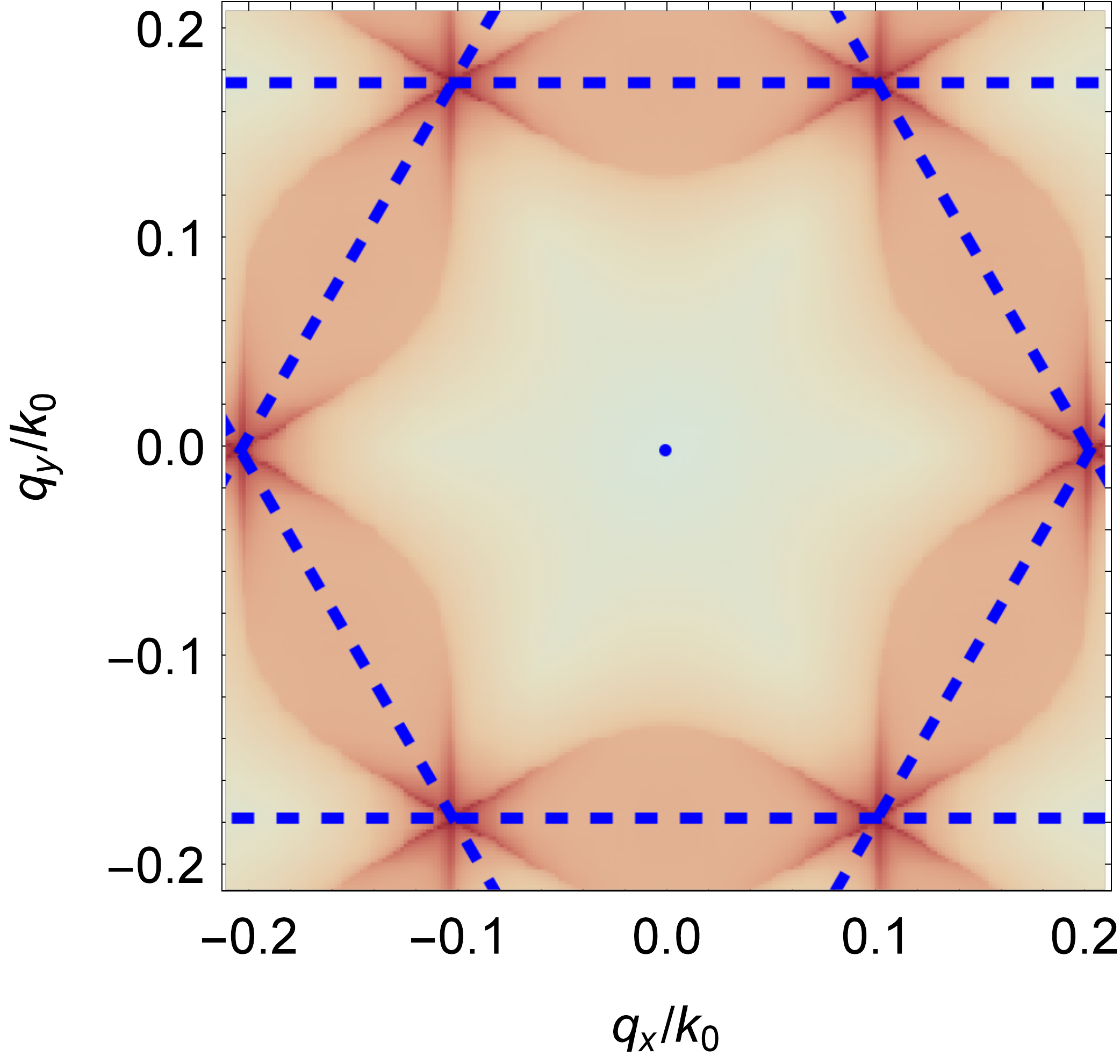}}\raisebox{\raiselength\height}{\includegraphics[height=\legendheight\textwidth]{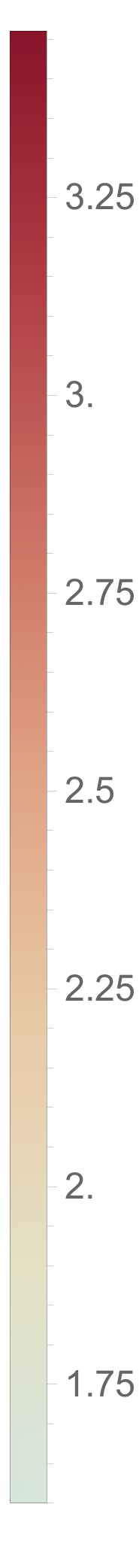}}
    \caption{\label{figJLargestEVLatticeHexContraAligned0p577}$a= \frac{2 \pi}{\sqrt{3}}Q^{-1}$}
  \end{subfigure}
  \begin{subfigure}{\smallLength}
    \raisebox{-0.5 \height}{\includegraphics[width=\graphicwidth \textwidth]{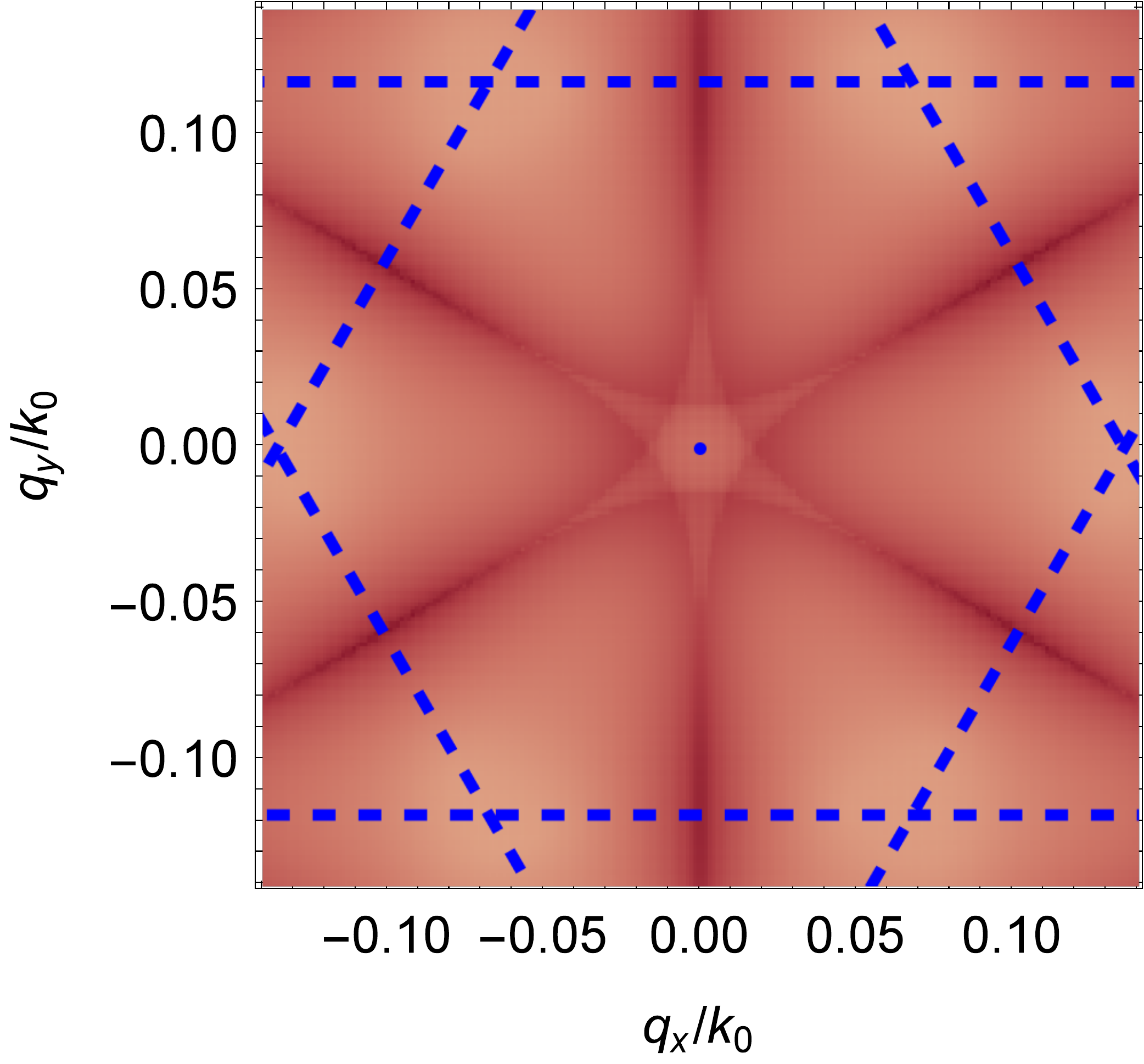}}\raisebox{\raiselength\height}{\includegraphics[height=\legendheight\textwidth]{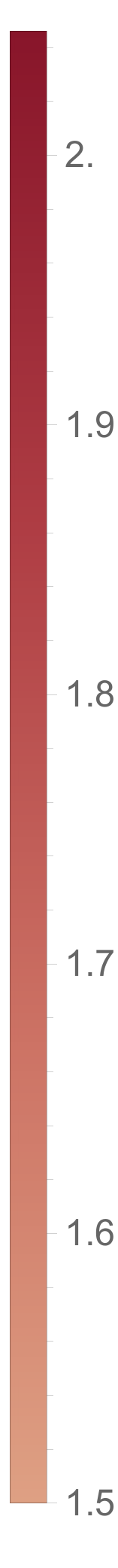}}
    \caption{\label{figJLargestEVLatticeHexContraAligned0p866}$a= \frac{3 \pi}{\sqrt{3}}Q^{-1}$}
  \end{subfigure}

\caption{\label{figJForDifferentLattices}(Color online) The largest eigenvalue of the RKKY interaction in momentum space for different lattice structures and lattice constants, $a$. The borders of the Brillouin zones are marked by dotted lines. A square, hexagonal, and $\pi/3$ rotated hexagonal lattice, is shown in the top, middle, and bottom panels, respectively.
}
\end{figure}

\subsection{\label{E} E. Spiral ground state and Temperature Stability}
In this section, we give details on the Monte Carlo algorithm used to determine the transition temperature to a spiral ground state.
To facilitate the numerics, we approximate $J$ by its contributions close the peaks $P$, i.e.,
\begin{align}
\hat{J}^{\lambda,\lambda^\prime}_p = \sum_{q \in \pm\{Q_i\}_{i=1}^3} 
\mbox{rect} \left(\frac{k_x-q_x}{\sqrt{A}} \right) 
\mbox{rect} \left(\frac{k_y-q_y}{\sqrt{A}} \right)
J^{\lambda,\lambda^\prime}_q,
\label{eqApproximationOfJ}
\end{align}
where $A$ approximates the area of the peak in momentum space and $\mbox{rect}$ is the rectangular function. From \refFig{figSpinSusceptibilityPPContribution}, we estimate $A \approx 0.1 k_0^2$.
A broad class of systems with a J-matrix consisting of a finite number of peaks is expected to develop a spin wave in their ground states.
We numerically determine the energetically minimal spin configurations of a $16 \times 16$ periodic hexagonal lattice of spins with $a = \frac{\pi}{4}Q^{-1}$.
Starting from a random spin distribution where every configuration is equally likely, we thermalize the system using a Monte Carlo algorithm\cite{Metropolis1953}. For $\beta \rightarrow \infty$, we reach local energetic minima.
We find that the initial configurations relax to one of three spiral spin waves with momenta $Q_1$, $Q_2$, and $Q_3$, cf. \refFig{figHexagonalFermiSurfaceRange}. One of them is shown in Fig.~5a in the main text. The remaining two ground state configurations are related to the one in Fig.~5a by a rotation around the $z$ axis about an angle of $\pm\pi/3$.
Note, that each spiral wave has a phase degeneracy, equivalent to shifting the origin of the spiral.

\subsubsection{Determination of the transition temperature}
The main text introduces the material parameter $\zeta = \frac{E_0}{k_0^2}$, and the inverse transition temperature is given by $\beta_c \approx 2.4 \alpha \zeta \tilde{\beta}_c = 2.4 * 2\pi \frac{V_{\text{uc}} E_0}{J_0^2 k_0^2} \tilde{\beta}_c$.
We estimate values of the exchange coupling $J_0$ in order of magnitude approximation by assigning an exchange energy of the order of $0.1 -1 \mathrm{eV}$ per unit cell of the topological insulator. This yields $J \approx 10 - 10^2 \mathrm{meV} \mathrm{nm}^2$. We consider a lattice constant of $a = 1 \mathrm{nm}$. Furthermore, we 
allow for $k_0^2 / E_0$ ranging from $1.5 \mathrm{eV}^{-1} \mathrm{nm}^{-2}$ for Bi$_2$Te$_2$Se to $3.8 \mathrm{eV}^{-1} \mathrm{nm}^{-2}$ for Bi$_2$Te$_3$, which we take from the table in \citeRef{JiangWu2011SpinSusceptibilityAndHelicalMagneticOrderAtTheEdgesSurfacesOfTopologicalInsulatorsDueToFermiSurfaceNesting}.
With these values, the critical temperature $T_c$  lies between $0.12 K$ and $30 K$ corresponding to the critical temperature presented in the main text.


\begin{thebibliography}{39}%
\makeatletter
\providecommand \@ifxundefined [1]{%
 \@ifx{#1\undefined}
}%
\providecommand \@ifnum [1]{%
 \ifnum #1\expandafter \@firstoftwo
 \else \expandafter \@secondoftwo
 \fi
}%
\providecommand \@ifx [1]{%
 \ifx #1\expandafter \@firstoftwo
 \else \expandafter \@secondoftwo
 \fi
}%
\providecommand \natexlab [1]{#1}%
\providecommand \enquote  [1]{``#1''}%
\providecommand \bibnamefont  [1]{#1}%
\providecommand \bibfnamefont [1]{#1}%
\providecommand \citenamefont [1]{#1}%
\providecommand \href@noop [0]{\@secondoftwo}%
\providecommand \href [0]{\begingroup \@sanitize@url \@href}%
\providecommand \@href[1]{\@@startlink{#1}\@@href}%
\providecommand \@@href[1]{\endgroup#1\@@endlink}%
\providecommand \@sanitize@url [0]{\catcode `\\12\catcode `\$12\catcode
  `\&12\catcode `\#12\catcode `\^12\catcode `\_12\catcode `\%12\relax}%
\providecommand \@@startlink[1]{}%
\providecommand \@@endlink[0]{}%
\providecommand \url  [0]{\begingroup\@sanitize@url \@url }%
\providecommand \@url [1]{\endgroup\@href {#1}{\urlprefix }}%
\providecommand \urlprefix  [0]{URL }%
\providecommand \Eprint [0]{\href }%
\providecommand \doibase [0]{http://dx.doi.org/}%
\providecommand \selectlanguage [0]{\@gobble}%
\providecommand \bibinfo  [0]{\@secondoftwo}%
\providecommand \bibfield  [0]{\@secondoftwo}%
\providecommand \translation [1]{[#1]}%
\providecommand \BibitemOpen [0]{}%
\providecommand \bibitemStop [0]{}%
\providecommand \bibitemNoStop [0]{.\EOS\space}%
\providecommand \EOS [0]{\spacefactor3000\relax}%
\providecommand \BibitemShut  [1]{\csname bibitem#1\endcsname}%
\let\auto@bib@innerbib\@empty
\bibitem [{\citenamefont {Hasan}\ and\ \citenamefont {Kane}(2010)}]{Hasan2010}%
  \BibitemOpen
  \bibfield  {author} {\bibinfo {author} {\bibfnamefont {M.}~\bibnamefont
  {Hasan}}\ and\ \bibinfo {author} {\bibfnamefont {C.}~\bibnamefont {Kane}},\
  }\href {\doibase 10.1103/RevModPhys.82.3045} {\bibfield  {journal} {\bibinfo
  {journal} {Rev. Mod. Phys.}\ }\textbf {\bibinfo {volume} {82}},\ \bibinfo
  {pages} {3045} (\bibinfo {year} {2010})}\BibitemShut {NoStop}%
\bibitem [{\citenamefont {Fu}\ and\ \citenamefont {Kane}(2008)}]{FuandKane}%
  \BibitemOpen
  \bibfield  {author} {\bibinfo {author} {\bibfnamefont {L.}~\bibnamefont
  {Fu}}\ and\ \bibinfo {author} {\bibfnamefont {C.~L.}\ \bibnamefont {Kane}},\
  }\href {\doibase 10.1103/PhysRevLett.100.096407} {\bibfield  {journal}
  {\bibinfo  {journal} {Phys. Rev. Lett.}\ }\textbf {\bibinfo {volume} {100}},\
  \bibinfo {pages} {096407} (\bibinfo {year} {2008})}\BibitemShut {NoStop}%
\bibitem [{\citenamefont {Stern}(2010)}]{nonabelian}%
  \BibitemOpen
  \bibfield  {author} {\bibinfo {author} {\bibfnamefont {A.}~\bibnamefont
  {Stern}},\ }\href {\doibase 10.1038/nature08915} {\bibfield  {journal}
  {\bibinfo  {journal} {Nature}\ }\textbf {\bibinfo {volume} {464}},\ \bibinfo
  {pages} {187} (\bibinfo {year} {2010})}\BibitemShut {NoStop}%
\bibitem [{\citenamefont {Deng}\ \emph {et~al.}(2014)\citenamefont {Deng},
  \citenamefont {Ortiz}, \citenamefont {Poudel},\ and\ \citenamefont
  {Viola}}]{Majoranaflatbands}%
  \BibitemOpen
  \bibfield  {author} {\bibinfo {author} {\bibfnamefont {S.}~\bibnamefont
  {Deng}}, \bibinfo {author} {\bibfnamefont {G.}~\bibnamefont {Ortiz}},
  \bibinfo {author} {\bibfnamefont {A.}~\bibnamefont {Poudel}}, \ and\ \bibinfo
  {author} {\bibfnamefont {L.}~\bibnamefont {Viola}},\ }\href {\doibase
  10.1103/PhysRevB.89.140507} {\bibfield  {journal} {\bibinfo  {journal} {Phys.
  Rev. B}\ }\textbf {\bibinfo {volume} {89}},\ \bibinfo {pages} {140507}
  (\bibinfo {year} {2014})}\BibitemShut {NoStop}%
\bibitem [{\citenamefont {{Keselman}}\ and\ \citenamefont
  {{Berg}}(2015)}]{Keselman}%
  \BibitemOpen
  \bibfield  {author} {\bibinfo {author} {\bibfnamefont {A.}~\bibnamefont
  {{Keselman}}}\ and\ \bibinfo {author} {\bibfnamefont {E.}~\bibnamefont
  {{Berg}}},\ }\href@noop {} {\bibfield  {journal} {\bibinfo  {journal} {ArXiv
  e-prints}\ } (\bibinfo {year} {2015})},\ \Eprint
  {http://arxiv.org/abs/1502.02037} {arXiv:1502.02037 [cond-mat.mes-hall]}
  \BibitemShut {NoStop}%
\bibitem [{\citenamefont {Matsuura}\ \emph {et~al.}(2013)\citenamefont
  {Matsuura}, \citenamefont {Chang}, \citenamefont {Schnyder},\ and\
  \citenamefont {Ryu}}]{Schnyder2013}%
  \BibitemOpen
  \bibfield  {author} {\bibinfo {author} {\bibfnamefont {S.}~\bibnamefont
  {Matsuura}}, \bibinfo {author} {\bibfnamefont {P.-Y.}\ \bibnamefont {Chang}},
  \bibinfo {author} {\bibfnamefont {A.~P.}\ \bibnamefont {Schnyder}}, \ and\
  \bibinfo {author} {\bibfnamefont {S.}~\bibnamefont {Ryu}},\ }\href
  {http://stacks.iop.org/1367-2630/15/i=6/a=065001} {\bibfield  {journal}
  {\bibinfo  {journal} {New Journal of Physics}\ }\textbf {\bibinfo {volume}
  {15}},\ \bibinfo {pages} {065001} (\bibinfo {year} {2013})}\BibitemShut
  {NoStop}%
\bibitem [{\citenamefont {Queiroz}\ and\ \citenamefont
  {Schnyder}(2014)}]{Queiroz}%
  \BibitemOpen
  \bibfield  {author} {\bibinfo {author} {\bibfnamefont {R.}~\bibnamefont
  {Queiroz}}\ and\ \bibinfo {author} {\bibfnamefont {A.~P.}\ \bibnamefont
  {Schnyder}},\ }\href {\doibase 10.1103/PhysRevB.89.054501} {\bibfield
  {journal} {\bibinfo  {journal} {Phys. Rev. B}\ }\textbf {\bibinfo {volume}
  {89}},\ \bibinfo {pages} {054501} (\bibinfo {year} {2014})}\BibitemShut
  {NoStop}%
\bibitem [{\citenamefont {{Baum}}\ \emph {et~al.}(2014)\citenamefont {{Baum}},
  \citenamefont {{Posske}}, \citenamefont {{Cosma Fulga}}, \citenamefont
  {{Trauzettel}},\ and\ \citenamefont {{Stern}}}]{ourpaper}%
  \BibitemOpen
  \bibfield  {author} {\bibinfo {author} {\bibfnamefont {Y.}~\bibnamefont
  {{Baum}}}, \bibinfo {author} {\bibfnamefont {T.}~\bibnamefont {{Posske}}},
  \bibinfo {author} {\bibfnamefont {I.}~\bibnamefont {{Cosma Fulga}}}, \bibinfo
  {author} {\bibfnamefont {B.}~\bibnamefont {{Trauzettel}}}, \ and\ \bibinfo
  {author} {\bibfnamefont {A.}~\bibnamefont {{Stern}}},\ }\href@noop {}
  {\bibfield  {journal} {\bibinfo  {journal} {ArXiv e-prints}\ } (\bibinfo
  {year} {2014})},\ \Eprint {http://arxiv.org/abs/1412.0021} {arXiv:1412.0021
  [cond-mat.mes-hall]} \BibitemShut {NoStop}%
\bibitem [{\citenamefont {Schnyder}\ \emph {et~al.}(2008)\citenamefont
  {Schnyder}, \citenamefont {Ryu}, \citenamefont {Furusaki},\ and\
  \citenamefont {Ludwig}}]{class}%
  \BibitemOpen
  \bibfield  {author} {\bibinfo {author} {\bibfnamefont {A.~P.}\ \bibnamefont
  {Schnyder}}, \bibinfo {author} {\bibfnamefont {S.}~\bibnamefont {Ryu}},
  \bibinfo {author} {\bibfnamefont {A.}~\bibnamefont {Furusaki}}, \ and\
  \bibinfo {author} {\bibfnamefont {A.~W.~W.}\ \bibnamefont {Ludwig}},\ }\href
  {\doibase 10.1103/PhysRevB.78.195125} {\bibfield  {journal} {\bibinfo
  {journal} {Phys. Rev. B}\ }\textbf {\bibinfo {volume} {78}},\ \bibinfo
  {pages} {195125} (\bibinfo {year} {2008})}\BibitemShut {NoStop}%
\bibitem [{\citenamefont {{Kitaev}}(2009)}]{Kitaev}%
  \BibitemOpen
  \bibfield  {author} {\bibinfo {author} {\bibfnamefont {A.}~\bibnamefont
  {{Kitaev}}},\ }in\ \href {\doibase 10.1063/1.3149495} {\emph {\bibinfo
  {booktitle} {American Institute of Physics Conference Series}}},\ \bibinfo
  {series} {American Institute of Physics Conference Series}, Vol.\ \bibinfo
  {volume} {1134},\ \bibinfo {editor} {edited by\ \bibinfo {editor}
  {\bibfnamefont {V.}~\bibnamefont {{Lebedev}}}\ and\ \bibinfo {editor}
  {\bibfnamefont {M.}~\bibnamefont {{Feigel'man}}}}\ (\bibinfo {year} {2009})\
  pp.\ \bibinfo {pages} {22--30}\BibitemShut {NoStop}%
\bibitem [{SCp()}]{SCparameters}%
  \BibitemOpen
  \href@noop {} {}\bibinfo {note} {The calculation is done in the 2-terminal
  setup with periodic boundary conditions. The lattice size is $80\times 80$,
  and the parameters are $m=0.2$ and $\Delta=0.14$. Each point is obtained by
  averaging over $50$ independent realizations of disorder.}\BibitemShut
  {Stop}%
\bibitem [{Note1()}]{Note1}%
  \BibitemOpen
  \bibinfo {note} {For the practical calculation, we assume that the magnetic
  periodicity is commensurate with the lattice periodicity}\BibitemShut
  {NoStop}%
\bibitem [{\citenamefont {Sau}\ \emph {et~al.}(2010)\citenamefont {Sau},
  \citenamefont {Lutchyn}, \citenamefont {Tewari},\ and\ \citenamefont
  {Das~Sarma}}]{Rashba1}%
  \BibitemOpen
  \bibfield  {author} {\bibinfo {author} {\bibfnamefont {J.~D.}\ \bibnamefont
  {Sau}}, \bibinfo {author} {\bibfnamefont {R.~M.}\ \bibnamefont {Lutchyn}},
  \bibinfo {author} {\bibfnamefont {S.}~\bibnamefont {Tewari}}, \ and\ \bibinfo
  {author} {\bibfnamefont {S.}~\bibnamefont {Das~Sarma}},\ }\href {\doibase
  10.1103/PhysRevLett.104.040502} {\bibfield  {journal} {\bibinfo  {journal}
  {Phys. Rev. Lett.}\ }\textbf {\bibinfo {volume} {104}},\ \bibinfo {pages}
  {040502} (\bibinfo {year} {2010})}\BibitemShut {NoStop}%
\bibitem [{\citenamefont {Liu}\ \emph {et~al.}(2009)\citenamefont {Liu},
  \citenamefont {Liu}, \citenamefont {Xu}, \citenamefont {Qi},\ and\
  \citenamefont
  {Zhang}}]{LiuLiuXuQiZhang2009MagneticImpuritiesOnTheSurfaceOfATI}%
  \BibitemOpen
  \bibfield  {author} {\bibinfo {author} {\bibfnamefont {Q.}~\bibnamefont
  {Liu}}, \bibinfo {author} {\bibfnamefont {C.-X.}\ \bibnamefont {Liu}},
  \bibinfo {author} {\bibfnamefont {C.}~\bibnamefont {Xu}}, \bibinfo {author}
  {\bibfnamefont {X.-L.}\ \bibnamefont {Qi}}, \ and\ \bibinfo {author}
  {\bibfnamefont {S.-C.}\ \bibnamefont {Zhang}},\ }\href {\doibase
  10.1103/PhysRevLett.102.156603} {\bibfield  {journal} {\bibinfo  {journal}
  {Phys. Rev. Lett.}\ }\textbf {\bibinfo {volume} {102}},\ \bibinfo {pages}
  {156603} (\bibinfo {year} {2009})}\BibitemShut {NoStop}%
\bibitem [{\citenamefont {Schmidt}\ \emph {et~al.}(2011)\citenamefont
  {Schmidt}, \citenamefont {Miwa},\ and\ \citenamefont
  {Fazzio}}]{SchmidtMiwaFazzio2011SpinTextureAndMagneticAnisotropyOfCoImpuritiesInBi2Se3TI}%
  \BibitemOpen
  \bibfield  {author} {\bibinfo {author} {\bibfnamefont {T.~M.}\ \bibnamefont
  {Schmidt}}, \bibinfo {author} {\bibfnamefont {R.~H.}\ \bibnamefont {Miwa}}, \
  and\ \bibinfo {author} {\bibfnamefont {A.}~\bibnamefont {Fazzio}},\ }\href
  {\doibase 10.1103/PhysRevB.84.245418} {\bibfield  {journal} {\bibinfo
  {journal} {Phys. Rev. B}\ }\textbf {\bibinfo {volume} {84}},\ \bibinfo
  {pages} {245418} (\bibinfo {year} {2011})}\BibitemShut {NoStop}%
\bibitem [{\citenamefont {Rosenberg}\ and\ \citenamefont
  {Franz}(2012)}]{RosenbergFranz2012SurfaceMagneticOrderingInTopologicalInsulatorsWithBulkMagneticDopants}%
  \BibitemOpen
  \bibfield  {author} {\bibinfo {author} {\bibfnamefont {G.}~\bibnamefont
  {Rosenberg}}\ and\ \bibinfo {author} {\bibfnamefont {M.}~\bibnamefont
  {Franz}},\ }\href {\doibase 10.1103/PhysRevB.85.195119} {\bibfield  {journal}
  {\bibinfo  {journal} {Phys. Rev. B}\ }\textbf {\bibinfo {volume} {85}},\
  \bibinfo {pages} {195119} (\bibinfo {year} {2012})}\BibitemShut {NoStop}%
\bibitem [{\citenamefont {Caprara}\ \emph {et~al.}(2012)\citenamefont
  {Caprara}, \citenamefont {Tugushev}, \citenamefont {Echenique},\ and\
  \citenamefont
  {Chulkov}}]{CapraraTugushevEcheniqueChulkov2012SpinPolarizedStatesOfMatterOnTheSurfaceOfA3DTIWithImplantedMagneticAtoms}%
  \BibitemOpen
  \bibfield  {author} {\bibinfo {author} {\bibfnamefont {S.}~\bibnamefont
  {Caprara}}, \bibinfo {author} {\bibfnamefont {V.~V.}\ \bibnamefont
  {Tugushev}}, \bibinfo {author} {\bibfnamefont {P.~M.}\ \bibnamefont
  {Echenique}}, \ and\ \bibinfo {author} {\bibfnamefont {E.~V.}\ \bibnamefont
  {Chulkov}},\ }\href {\doibase 10.1103/PhysRevB.85.121304} {\bibfield
  {journal} {\bibinfo  {journal} {Phys. Rev. B}\ }\textbf {\bibinfo {volume}
  {85}},\ \bibinfo {pages} {121304} (\bibinfo {year} {2012})}\BibitemShut
  {NoStop}%
\bibitem [{\citenamefont {Zhu}\ \emph {et~al.}(2011)\citenamefont {Zhu},
  \citenamefont {Yao}, \citenamefont {Zhang},\ and\ \citenamefont
  {Chang}}]{ZhuYaoZhangChang2011ElectricallyControllableSurfaceMagnetismOnTheSurfaceOfTopologicalInsulators}%
  \BibitemOpen
  \bibfield  {author} {\bibinfo {author} {\bibfnamefont {J.-J.}\ \bibnamefont
  {Zhu}}, \bibinfo {author} {\bibfnamefont {D.-X.}\ \bibnamefont {Yao}},
  \bibinfo {author} {\bibfnamefont {S.-C.}\ \bibnamefont {Zhang}}, \ and\
  \bibinfo {author} {\bibfnamefont {K.}~\bibnamefont {Chang}},\ }\href
  {\doibase 10.1103/PhysRevLett.106.097201} {\bibfield  {journal} {\bibinfo
  {journal} {Phys. Rev. Lett.}\ }\textbf {\bibinfo {volume} {106}},\ \bibinfo
  {pages} {097201} (\bibinfo {year} {2011})}\BibitemShut {NoStop}%
\bibitem [{\citenamefont {{Ye}}\ \emph {et~al.}(2010)\citenamefont {{Ye}},
  \citenamefont {{Ding}}, \citenamefont {{Zhai}},\ and\ \citenamefont
  {{Su}}}]{YeDingZhaSu2010SpinHelixOfMagneticImpuritiesIn2DHelicalMetal}%
  \BibitemOpen
  \bibfield  {author} {\bibinfo {author} {\bibfnamefont {F.}~\bibnamefont
  {{Ye}}}, \bibinfo {author} {\bibfnamefont {G.~H.}\ \bibnamefont {{Ding}}},
  \bibinfo {author} {\bibfnamefont {H.}~\bibnamefont {{Zhai}}}, \ and\ \bibinfo
  {author} {\bibfnamefont {Z.~B.}\ \bibnamefont {{Su}}},\ }\href {\doibase
  10.1209/0295-5075/90/47001} {\bibfield  {journal} {\bibinfo  {journal} {EPL
  (Europhysics Letters)}\ }\textbf {\bibinfo {volume} {90}},\ \bibinfo {pages}
  {47001} (\bibinfo {year} {2010})}\BibitemShut {NoStop}%
\bibitem [{\citenamefont {Klinovaja}\ \emph {et~al.}(2013)\citenamefont
  {Klinovaja}, \citenamefont {Stano}, \citenamefont {Yazdani},\ and\
  \citenamefont
  {Loss}}]{KlinovajaStanoYazdaniLoss2013TopologicalSuperconductivityAndMajoranaFermionsInRKKYSystems}%
  \BibitemOpen
  \bibfield  {author} {\bibinfo {author} {\bibfnamefont {J.}~\bibnamefont
  {Klinovaja}}, \bibinfo {author} {\bibfnamefont {P.}~\bibnamefont {Stano}},
  \bibinfo {author} {\bibfnamefont {A.}~\bibnamefont {Yazdani}}, \ and\
  \bibinfo {author} {\bibfnamefont {D.}~\bibnamefont {Loss}},\ }\href {\doibase
  10.1103/PhysRevLett.111.186805} {\bibfield  {journal} {\bibinfo  {journal}
  {Phys. Rev. Lett.}\ }\textbf {\bibinfo {volume} {111}},\ \bibinfo {pages}
  {186805} (\bibinfo {year} {2013})}\BibitemShut {NoStop}%
\bibitem [{\citenamefont {Jiang}\ and\ \citenamefont
  {Wu}(2011)}]{JiangWu2011SpinSusceptibilityAndHelicalMagneticOrderAtTheEdgesSurfacesOfTopologicalInsulatorsDueToFermiSurfaceNesting}%
  \BibitemOpen
  \bibfield  {author} {\bibinfo {author} {\bibfnamefont {J.-H.}\ \bibnamefont
  {Jiang}}\ and\ \bibinfo {author} {\bibfnamefont {S.}~\bibnamefont {Wu}},\
  }\href {\doibase 10.1103/PhysRevB.83.205124} {\bibfield  {journal} {\bibinfo
  {journal} {Phys. Rev. B}\ }\textbf {\bibinfo {volume} {83}},\ \bibinfo
  {pages} {205124} (\bibinfo {year} {2011})}\BibitemShut {NoStop}%
\bibitem [{\citenamefont {{Zyuzin}}\ and\ \citenamefont
  {{Loss}}(2014)}]{LossRKKYpaper}%
  \BibitemOpen
  \bibfield  {author} {\bibinfo {author} {\bibfnamefont {A.~A.}\ \bibnamefont
  {{Zyuzin}}}\ and\ \bibinfo {author} {\bibfnamefont {D.}~\bibnamefont
  {{Loss}}},\ }\href@noop {} {\bibfield  {journal} {\bibinfo  {journal} {ArXiv
  e-prints}\ } (\bibinfo {year} {2014})},\ \Eprint
  {http://arxiv.org/abs/1407.6632} {arXiv:1407.6632 [cond-mat.mes-hall]}
  \BibitemShut {NoStop}%
\bibitem [{\citenamefont {{Zahid Hasan}}\ \emph {et~al.}(2014)\citenamefont
  {{Zahid Hasan}}, \citenamefont {{Xu}}, \citenamefont {{Hsieh}}, \citenamefont
  {{Wray}},\ and\ \citenamefont {{Xia}}}]{Hasan2014}%
  \BibitemOpen
  \bibfield  {author} {\bibinfo {author} {\bibfnamefont {M.}~\bibnamefont
  {{Zahid Hasan}}}, \bibinfo {author} {\bibfnamefont {S.-Y.}\ \bibnamefont
  {{Xu}}}, \bibinfo {author} {\bibfnamefont {D.}~\bibnamefont {{Hsieh}}},
  \bibinfo {author} {\bibfnamefont {L.~A.}\ \bibnamefont {{Wray}}}, \ and\
  \bibinfo {author} {\bibfnamefont {Y.}~\bibnamefont {{Xia}}},\ }\href@noop {}
  {\bibfield  {journal} {\bibinfo  {journal} {ArXiv e-prints}\ } (\bibinfo
  {year} {2014})},\ \Eprint {http://arxiv.org/abs/1401.0848} {arXiv:1401.0848
  [cond-mat.mes-hall]} \BibitemShut {NoStop}%
\bibitem [{\citenamefont {Neupane}\ \emph {et~al.}(2012)\citenamefont
  {Neupane}, \citenamefont {Xu}, \citenamefont {Wray}, \citenamefont
  {Petersen}, \citenamefont {Shankar}, \citenamefont {Alidoust}, \citenamefont
  {Liu}, \citenamefont {Fedorov}, \citenamefont {Ji}, \citenamefont {Allred},
  \citenamefont {Hor}, \citenamefont {Chang}, \citenamefont {Jeng},
  \citenamefont {Lin}, \citenamefont {Bansil}, \citenamefont {Cava},\ and\
  \citenamefont {Hasan}}]{Neupane2012_several_ternary_materials}%
  \BibitemOpen
  \bibfield  {author} {\bibinfo {author} {\bibfnamefont {M.}~\bibnamefont
  {Neupane}}, \bibinfo {author} {\bibfnamefont {S.-Y.}\ \bibnamefont {Xu}},
  \bibinfo {author} {\bibfnamefont {L.~A.}\ \bibnamefont {Wray}}, \bibinfo
  {author} {\bibfnamefont {A.}~\bibnamefont {Petersen}}, \bibinfo {author}
  {\bibfnamefont {R.}~\bibnamefont {Shankar}}, \bibinfo {author} {\bibfnamefont
  {N.}~\bibnamefont {Alidoust}}, \bibinfo {author} {\bibfnamefont
  {C.}~\bibnamefont {Liu}}, \bibinfo {author} {\bibfnamefont {A.}~\bibnamefont
  {Fedorov}}, \bibinfo {author} {\bibfnamefont {H.}~\bibnamefont {Ji}},
  \bibinfo {author} {\bibfnamefont {J.~M.}\ \bibnamefont {Allred}}, \bibinfo
  {author} {\bibfnamefont {Y.~S.}\ \bibnamefont {Hor}}, \bibinfo {author}
  {\bibfnamefont {T.-R.}\ \bibnamefont {Chang}}, \bibinfo {author}
  {\bibfnamefont {H.-T.}\ \bibnamefont {Jeng}}, \bibinfo {author}
  {\bibfnamefont {H.}~\bibnamefont {Lin}}, \bibinfo {author} {\bibfnamefont
  {A.}~\bibnamefont {Bansil}}, \bibinfo {author} {\bibfnamefont {R.~J.}\
  \bibnamefont {Cava}}, \ and\ \bibinfo {author} {\bibfnamefont {M.~Z.}\
  \bibnamefont {Hasan}},\ }\href {\doibase 10.1103/PhysRevB.85.235406}
  {\bibfield  {journal} {\bibinfo  {journal} {Phys. Rev. B}\ }\textbf {\bibinfo
  {volume} {85}},\ \bibinfo {pages} {235406} (\bibinfo {year}
  {2012})}\BibitemShut {NoStop}%
\bibitem [{\citenamefont {Miyamoto}\ \emph {et~al.}(2012)\citenamefont
  {Miyamoto}, \citenamefont {Kimura}, \citenamefont {Okuda}, \citenamefont
  {Miyahara}, \citenamefont {Kuroda}, \citenamefont {Namatame}, \citenamefont
  {Taniguchi}, \citenamefont {Eremeev}, \citenamefont {Menshchikova},
  \citenamefont {Chulkov}, \citenamefont {Kokh},\ and\ \citenamefont
  {Tereshchenko}}]{Miyamoto2012}%
  \BibitemOpen
  \bibfield  {author} {\bibinfo {author} {\bibfnamefont {K.}~\bibnamefont
  {Miyamoto}}, \bibinfo {author} {\bibfnamefont {A.}~\bibnamefont {Kimura}},
  \bibinfo {author} {\bibfnamefont {T.}~\bibnamefont {Okuda}}, \bibinfo
  {author} {\bibfnamefont {H.}~\bibnamefont {Miyahara}}, \bibinfo {author}
  {\bibfnamefont {K.}~\bibnamefont {Kuroda}}, \bibinfo {author} {\bibfnamefont
  {H.}~\bibnamefont {Namatame}}, \bibinfo {author} {\bibfnamefont
  {M.}~\bibnamefont {Taniguchi}}, \bibinfo {author} {\bibfnamefont {S.~V.}\
  \bibnamefont {Eremeev}}, \bibinfo {author} {\bibfnamefont {T.~V.}\
  \bibnamefont {Menshchikova}}, \bibinfo {author} {\bibfnamefont {E.~V.}\
  \bibnamefont {Chulkov}}, \bibinfo {author} {\bibfnamefont {K.~A.}\
  \bibnamefont {Kokh}}, \ and\ \bibinfo {author} {\bibfnamefont {O.~E.}\
  \bibnamefont {Tereshchenko}},\ }\href {\doibase
  10.1103/PhysRevLett.109.166802} {\bibfield  {journal} {\bibinfo  {journal}
  {Phys. Rev. Lett.}\ }\textbf {\bibinfo {volume} {109}},\ \bibinfo {pages}
  {166802} (\bibinfo {year} {2012})}\BibitemShut {NoStop}%
\bibitem [{\citenamefont {Sato}\ \emph {et~al.}(2010)\citenamefont {Sato},
  \citenamefont {Segawa}, \citenamefont {Guo}, \citenamefont {Sugawara},
  \citenamefont {Souma}, \citenamefont {Takahashi},\ and\ \citenamefont
  {Ando}}]{Sato2010}%
  \BibitemOpen
  \bibfield  {author} {\bibinfo {author} {\bibfnamefont {T.}~\bibnamefont
  {Sato}}, \bibinfo {author} {\bibfnamefont {K.}~\bibnamefont {Segawa}},
  \bibinfo {author} {\bibfnamefont {H.}~\bibnamefont {Guo}}, \bibinfo {author}
  {\bibfnamefont {K.}~\bibnamefont {Sugawara}}, \bibinfo {author}
  {\bibfnamefont {S.}~\bibnamefont {Souma}}, \bibinfo {author} {\bibfnamefont
  {T.}~\bibnamefont {Takahashi}}, \ and\ \bibinfo {author} {\bibfnamefont
  {Y.}~\bibnamefont {Ando}},\ }\href {\doibase 10.1103/PhysRevLett.105.136802}
  {\bibfield  {journal} {\bibinfo  {journal} {Phys. Rev. Lett.}\ }\textbf
  {\bibinfo {volume} {105}},\ \bibinfo {pages} {136802} (\bibinfo {year}
  {2010})}\BibitemShut {NoStop}%
\bibitem [{\citenamefont
  {Fu}(2009)}]{Fu2009HexagonalWarpingEffectsInTheSurfaceStatesOfTheTopologicalInsulatorBi2Te3}%
  \BibitemOpen
  \bibfield  {author} {\bibinfo {author} {\bibfnamefont {L.}~\bibnamefont
  {Fu}},\ }\href {\doibase 10.1103/PhysRevLett.103.266801} {\bibfield
  {journal} {\bibinfo  {journal} {Phys. Rev. Lett.}\ }\textbf {\bibinfo
  {volume} {103}},\ \bibinfo {pages} {266801} (\bibinfo {year}
  {2009})}\BibitemShut {NoStop}%
\bibitem [{\citenamefont {Ruderman}\ and\ \citenamefont
  {Kittel}(1954)}]{RudermanKittel1954IndirectExchangeCouplingOfNuclearMagneticMomentsByConductionElectrons}%
  \BibitemOpen
  \bibfield  {author} {\bibinfo {author} {\bibfnamefont {M.~A.}\ \bibnamefont
  {Ruderman}}\ and\ \bibinfo {author} {\bibfnamefont {C.}~\bibnamefont
  {Kittel}},\ }\href {\doibase 10.1103/PhysRev.96.99} {\bibfield  {journal}
  {\bibinfo  {journal} {Phys. Rev.}\ }\textbf {\bibinfo {volume} {96}},\
  \bibinfo {pages} {99} (\bibinfo {year} {1954})}\BibitemShut {NoStop}%
\bibitem [{\citenamefont
  {Kasuya}(1956)}]{Kasuya1956ATheoryOfMetallicFerroAndAntiferromagnetismOnZenersModel}%
  \BibitemOpen
  \bibfield  {author} {\bibinfo {author} {\bibfnamefont {T.}~\bibnamefont
  {Kasuya}},\ }\href {\doibase 10.1143/ptp.16.45} {\bibfield  {journal}
  {\bibinfo  {journal} {Progress of Theoretical Physics}\ }\textbf {\bibinfo
  {volume} {16}},\ \bibinfo {pages} {45} (\bibinfo {year} {1956})}\BibitemShut
  {NoStop}%
\bibitem [{\citenamefont
  {Yosida}(1957)}]{Yosida1957MagneticPropertiesOfCuMnAlloys}%
  \BibitemOpen
  \bibfield  {author} {\bibinfo {author} {\bibfnamefont {K.}~\bibnamefont
  {Yosida}},\ }\href {\doibase 10.1103/PhysRev.106.893} {\bibfield  {journal}
  {\bibinfo  {journal} {Phys. Rev.}\ }\textbf {\bibinfo {volume} {106}},\
  \bibinfo {pages} {893} (\bibinfo {year} {1957})}\BibitemShut {NoStop}%
\bibitem [{\citenamefont {Roth}\ \emph {et~al.}(1966)\citenamefont {Roth},
  \citenamefont {Zeiger},\ and\ \citenamefont
  {Kaplan}}]{RothZeigerKaplan1966GeneralizationOfTheRKKYInteractionForNonsphericalFermiSurfaces}%
  \BibitemOpen
  \bibfield  {author} {\bibinfo {author} {\bibfnamefont {L.~M.}\ \bibnamefont
  {Roth}}, \bibinfo {author} {\bibfnamefont {H.~J.}\ \bibnamefont {Zeiger}}, \
  and\ \bibinfo {author} {\bibfnamefont {T.~A.}\ \bibnamefont {Kaplan}},\
  }\href {\doibase 10.1103/PhysRev.149.519} {\bibfield  {journal} {\bibinfo
  {journal} {Phys. Rev.}\ }\textbf {\bibinfo {volume} {149}},\ \bibinfo {pages}
  {519} (\bibinfo {year} {1966})}\BibitemShut {NoStop}%
\bibitem [{\citenamefont {Nakamura}\ \emph {et~al.}(1985)\citenamefont
  {Nakamura}, \citenamefont {Nakahara}, \citenamefont {Ohtomi},\ and\
  \citenamefont
  {Sugano}}]{Nakamura1985_HexagonalFermiSurface_MechanismOf77ReconstructionOnSi111AndGe111SnSurfaces}%
  \BibitemOpen
  \bibfield  {author} {\bibinfo {author} {\bibfnamefont {K.}~\bibnamefont
  {Nakamura}}, \bibinfo {author} {\bibfnamefont {Y.}~\bibnamefont {Nakahara}},
  \bibinfo {author} {\bibfnamefont {K.}~\bibnamefont {Ohtomi}}, \ and\ \bibinfo
  {author} {\bibfnamefont {S.}~\bibnamefont {Sugano}},\ }\href {\doibase
  http://dx.doi.org/10.1016/0039-6028(85)90516-3} {\bibfield  {journal}
  {\bibinfo  {journal} {Surface Science}\ }\textbf {\bibinfo {volume} {152-153,
  Part 2}},\ \bibinfo {pages} {1020} (\bibinfo {year} {1985})}\BibitemShut
  {NoStop}%
\bibitem [{\citenamefont {Baum}\ and\ \citenamefont
  {Stern}(2012)}]{BaumStern2012MagneticInstabilityOnTheSurfaceOfTopologicalInsulators}%
  \BibitemOpen
  \bibfield  {author} {\bibinfo {author} {\bibfnamefont {Y.}~\bibnamefont
  {Baum}}\ and\ \bibinfo {author} {\bibfnamefont {A.}~\bibnamefont {Stern}},\
  }\href {\doibase 10.1103/PhysRevB.85.121105} {\bibfield  {journal} {\bibinfo
  {journal} {Phys. Rev. B}\ }\textbf {\bibinfo {volume} {85}},\ \bibinfo
  {pages} {121105} (\bibinfo {year} {2012})}\BibitemShut {NoStop}%
\bibitem [{Note2()}]{Note2}%
  \BibitemOpen
  \bibinfo {note} {The choice of the cutoff in the calculation of $\chi $ has
  been omitted in previous publications, which we improve upon by a numerical
  justification in Appendix~C.}\BibitemShut {Stop}%
\bibitem [{\citenamefont {Metropolis}\ \emph {et~al.}(1953)\citenamefont
  {Metropolis}, \citenamefont {Rosenbluth}, \citenamefont {Rosenbluth},
  \citenamefont {Teller},\ and\ \citenamefont {Teller}}]{Metropolis1953}%
  \BibitemOpen
  \bibfield  {author} {\bibinfo {author} {\bibfnamefont {N.}~\bibnamefont
  {Metropolis}}, \bibinfo {author} {\bibfnamefont {A.~W.}\ \bibnamefont
  {Rosenbluth}}, \bibinfo {author} {\bibfnamefont {M.~N.}\ \bibnamefont
  {Rosenbluth}}, \bibinfo {author} {\bibfnamefont {A.~H.}\ \bibnamefont
  {Teller}}, \ and\ \bibinfo {author} {\bibfnamefont {E.}~\bibnamefont
  {Teller}},\ }\href {\doibase http://dx.doi.org/10.1063/1.1699114} {\bibfield
  {journal} {\bibinfo  {journal} {The Journal of Chemical Physics}\ }\textbf
  {\bibinfo {volume} {21}},\ \bibinfo {pages} {1087} (\bibinfo {year}
  {1953})}\BibitemShut {NoStop}%
\bibitem [{Note3()}]{Note3}%
  \BibitemOpen
  \bibinfo {note} {For a finite number of simulated spins, the lower bound to
  the order parameter is $\rho = 12 \pi ^2/M^4$}\BibitemShut {NoStop}%
\bibitem [{\citenamefont {Polini}\ \emph {et~al.}(2013)\citenamefont {Polini},
  \citenamefont {Guinea}, \citenamefont {Lewenstein}, \citenamefont
  {Manoharan},\ and\ \citenamefont {Pellegrini}}]{Manoharan2013}%
  \BibitemOpen
  \bibfield  {author} {\bibinfo {author} {\bibfnamefont {M.}~\bibnamefont
  {Polini}}, \bibinfo {author} {\bibfnamefont {F.}~\bibnamefont {Guinea}},
  \bibinfo {author} {\bibfnamefont {M.}~\bibnamefont {Lewenstein}}, \bibinfo
  {author} {\bibfnamefont {H.~C.}\ \bibnamefont {Manoharan}}, \ and\ \bibinfo
  {author} {\bibfnamefont {V.}~\bibnamefont {Pellegrini}},\ }\href {\doibase
  10.1038/nnano.2013.161} {\bibfield  {journal} {\bibinfo  {journal} {Nat.
  Nanotechnol.}\ }\textbf {\bibinfo {volume} {8}},\ \bibinfo {pages} {625}
  (\bibinfo {year} {2013})}\BibitemShut {NoStop}%
\bibitem [{\citenamefont {Nadj-Perge}\ \emph {et~al.}(2014)\citenamefont
  {Nadj-Perge}, \citenamefont {Drozdov}, \citenamefont {Li}, \citenamefont
  {Chen}, \citenamefont {Jeon}, \citenamefont {Seo}, \citenamefont {MacDonald},
  \citenamefont {Bernevig},\ and\ \citenamefont {Yazdani}}]{Nadj-Perge2014}%
  \BibitemOpen
  \bibfield  {author} {\bibinfo {author} {\bibfnamefont {S.}~\bibnamefont
  {Nadj-Perge}}, \bibinfo {author} {\bibfnamefont {I.~K.}\ \bibnamefont
  {Drozdov}}, \bibinfo {author} {\bibfnamefont {J.}~\bibnamefont {Li}},
  \bibinfo {author} {\bibfnamefont {H.}~\bibnamefont {Chen}}, \bibinfo {author}
  {\bibfnamefont {S.}~\bibnamefont {Jeon}}, \bibinfo {author} {\bibfnamefont
  {J.}~\bibnamefont {Seo}}, \bibinfo {author} {\bibfnamefont {A.~H.}\
  \bibnamefont {MacDonald}}, \bibinfo {author} {\bibfnamefont {B.~A.}\
  \bibnamefont {Bernevig}}, \ and\ \bibinfo {author} {\bibfnamefont
  {A.}~\bibnamefont {Yazdani}},\ }\href {\doibase 10.1126/science.1259327}
  {\bibfield  {journal} {\bibinfo  {journal} {Science}\ }\textbf {\bibinfo
  {volume} {346}},\ \bibinfo {pages} {602} (\bibinfo {year} {2014})},\ \Eprint
  {http://arxiv.org/abs/http://www.sciencemag.org/content/346/6209/602.full.pdf}
  {http://www.sciencemag.org/content/346/6209/602.full.pdf} \BibitemShut
  {NoStop}%
\bibitem [{\citenamefont {Groth}\ \emph {et~al.}(2014)\citenamefont {Groth},
  \citenamefont {Wimmer}, \citenamefont {Akhmerov},\ and\ \citenamefont
  {Waintal}}]{kwant}%
  \BibitemOpen
  \bibfield  {author} {\bibinfo {author} {\bibfnamefont {C.~W.}\ \bibnamefont
  {Groth}}, \bibinfo {author} {\bibfnamefont {M.}~\bibnamefont {Wimmer}},
  \bibinfo {author} {\bibfnamefont {A.~R.}\ \bibnamefont {Akhmerov}}, \ and\
  \bibinfo {author} {\bibfnamefont {X.}~\bibnamefont {Waintal}},\ }\href
  {http://stacks.iop.org/1367-2630/16/i=6/a=063065} {\bibfield  {journal}
  {\bibinfo  {journal} {New Journal of Physics}\ }\textbf {\bibinfo {volume}
  {16}},\ \bibinfo {pages} {063065} (\bibinfo {year} {2014})}\BibitemShut
  {NoStop}%
\end{thebibliography}
%
\end{document}